\PassOptionsToPackage{table}{xcolor}
\documentclass[draft]{agujournal2019}
\usepackage{float}
\usepackage[table]{xcolor}
\usepackage{bbm}
\usepackage{graphicx,psfrag,amsmath,amsfonts,verbatim}
\usepackage{multirow}
\usepackage{tikz}
\usepackage{natbib}
\usepackage[caption=false]{subfig}
\usetikzlibrary{arrows.meta}
\tikzset{%
  >={Latex[width=2mm,length=2mm]},
            base/.style = {rectangle, rounded corners, draw=black,
                          minimum width=4cm, minimum height=1cm,
                          text centered, font=\sffamily},
              io/.style = {base, fill=blue!30},
      startstop/.style = {base, fill=red!30},
      decision/.style = {base, fill=green!30},
         process/.style = {base, minimum width=2.5cm, fill=orange!15, font=\ttfamily},
}
\usepackage{url} 
\usepackage{lineno}
\usepackage[inline]{trackchanges} 
\usepackage{soul}
\usepackage{mwe}
\draftfalse

\journalname{space weather}

\begin{document}

%
%


\title{Solar Flare Intensity Prediction with Machine Learning Models}

%
%




\authors{Zhenbang Jiao\affil{1}, Hu Sun\affil{1}, Xiantong Wang\affil{2}, Ward Manchester\affil{2}, Tamas Gombosi\affil{2}, Alfred Hero\affil{1,4}, Yang Chen\affil{1,3}}

 \affiliation{1}{Department of Statistics}
 \affiliation{2}{Climate and Space Sciences and Engineering}
 \affiliation{3}{Michigan Institute for Data Science}
 \affiliation{4}{Electrical Engineering and Computer Science}

\affiliation{1,2,3,4}{University of Michigan, Ann Arbor, MI, USA}




\correspondingauthor{Yang Chen}{ychenang@umich.edu}




\begin{keypoints}
\item We develop deep learning models to predict solar flare intensity values instead of flare classes from SHARP parameters in  {SDO/HMI} data set directly.
\item 
We use time-series information from both flaring time and non-flaring time in our model.
\item As opposed to solar flare classification, directly predicting solar flare intensity gives more detailed information about every occurrence of flares of each class. 
\end{keypoints}

%
%

%
%


\begin{abstract}
We develop a mixed Long Short Term Memory (LSTM) regression model to predict the maximum solar flare intensity within a 24-hour time window 0$\sim$24, 6$\sim$30, 12$\sim$36  {and} 24$\sim$48 hours ahead of time using 6, 12, 24  {and} 48 hours of data (predictors) for each  {Helioseismic and Magnetic Imager (HMI)} Active Region Patch (HARP). 
The model makes use of (1) the Space-weather HMI Active Region Patch (SHARP) parameters as predictors and (2) the exact flare intensities instead of class labels recorded in the Geostationary Operational Environmental Satellites (GOES) data set, which serves as the source of the response variables. Compared to solar flare classification, the model offers us more detailed information about the exact maximum flux level, i.e. intensity, for each occurrence of a flare. We also consider classification models built on top of the regression model and obtain better results in solar flare classifications as compared to \citet{doi:10.1029/2019SW002214}. Our results suggest that the most efficient time period for predicting the solar activity is within 24 hours before the prediction time using the SHARP parameters and the LSTM model.
\end{abstract}


%
%

%


\section{Introduction}

Space weather involves the dynamical processes of the Sun-Earth system that may affect human life and technology. The most destructive consequences of space weather, ranging from electric power disruptions to radiation hazards for astronauts, are due to energetic solar eruptions: producing both magnetic disturbances in the solar wind known as coronal mass ejections (CMEs) and intense electromagnetic radiation known as solar flares.   

Given their destructive capability, the predictions of energetic space weather events is critical for safeguarding our technological infrastructure. Extreme space storms -- those that could significantly degrade critical infrastructure -- could disable large portions of the electrical power grid, resulting in cascading failures that would affect key services such as water supply, health care, and transportation. The threat-assessment report by the Lloyd's insurance company \citep{Lloyds:2013_chip_a} concludes that extreme events could cause \$2.6 trillion in damage with a recovery time of months. An earlier report by the National Research Council \citep{Baker:2009a} arrived at similar conclusions.

While there are known precursors to these eruptions, accurate predictions of their occurrence remain very difficult. The current space weather forecasting based on physical models is far from reliable: the forecasting window is only minutes away from the current time point and the accuracy is low.  Previous work has established that solar eruptions are all associated with highly nonpotential magnetic fields that store the necessary free energy. The most energetic flares come from very localized intense kiloGauss photospheric fields known as active regions \citep{Forbes:2000a, Schrijver:2009a}. Measurement of these fields was greatly increased by the advent of  { the Helioseismic and Magnetic Imager (HMI) }   instrument on the Solar Dynamics Observatory (SDO) launched on February, 2010. HMI provides vast quantities of data in the form of high-cadence high-resolution vector magnetograms. These data are subdivided into HMI-Active Regions Patches (HARPs), which correspond to localized regions of intense magnetic fields. While HARPs are very similar to NOAA active regions they frequently define different spatial regions. Parameters relevant to solar eruptions are calculated from the HARP vector magnetic fields and saved with the data files which are designated as Space-weather HMI Active Region Patches, or SHARPs \citep{Bobra:2014a}.

Currently, over 7000 HARPs have been recorded, each one with full vector data saved on a 12-minute cadence for a period of approximately 14 days required to rotate across the disk. How to make the best use of the large amount of data available to provide reliable real-time forecasting of space weather events is one of the major questions for scientists in the field. Recently, data-driven approaches are gaining attention in the space science community with much more data becoming available. Scientists have adopted different machine learning algorithms to perform various space weather prediction tasks, including the solar flare classification using the  {SDO/HMI} SHARP parameters and other data sets, see~\citet{Barnes_2016}, ~\citet{Leka:2018b}, ~\citet{Liu_2019}, ~\citet{Camporeale:2019a}, ~\citet{Leka_2019_2} and ~\citet{Leka_2019_1} for reviews and references therein.  {Among all the papers mentioned, \citet{Liu_2019} also used the GOES data set and adopted the LSTM technique to predict solar flares. In contrast, in this paper we propose a different mixed LSTM model and we consider not only classification but also regression to predict the exact intensities rather than the labels of the solar flares. Moreover, our data pre-processing gives a new way of defining response variables and takes quiet time data into consideration. }

\citet{doi:10.1029/2019SW002214} showed that the time series of SHARP parameters from the  {SDO/HMI} data provide useful information for distinguishing strong solar flares of M/X class from weak flares of A/B class roughly 24 hours prior to the flare event. These SHARP parameters are derived from the HMI images based on physically meaningful quantities of the active regions where the flares emerge from, see~\citet{Bobra:2014a} for detailed descriptions of these features. To make the task of binary classification manageable, \citet{doi:10.1029/2019SW002214} only considered the B and M/X flares, ignoring the more prevalent C flares. This design is due to the consideration that flare classes are arbitrarily categorized based on a continuous logarithmic scale of flare intensity (radiant power level), thus strong C flares are essentially indistinguishable from weak M flares.

\begin{figure}
    \centering
    \includegraphics[width=\textwidth]{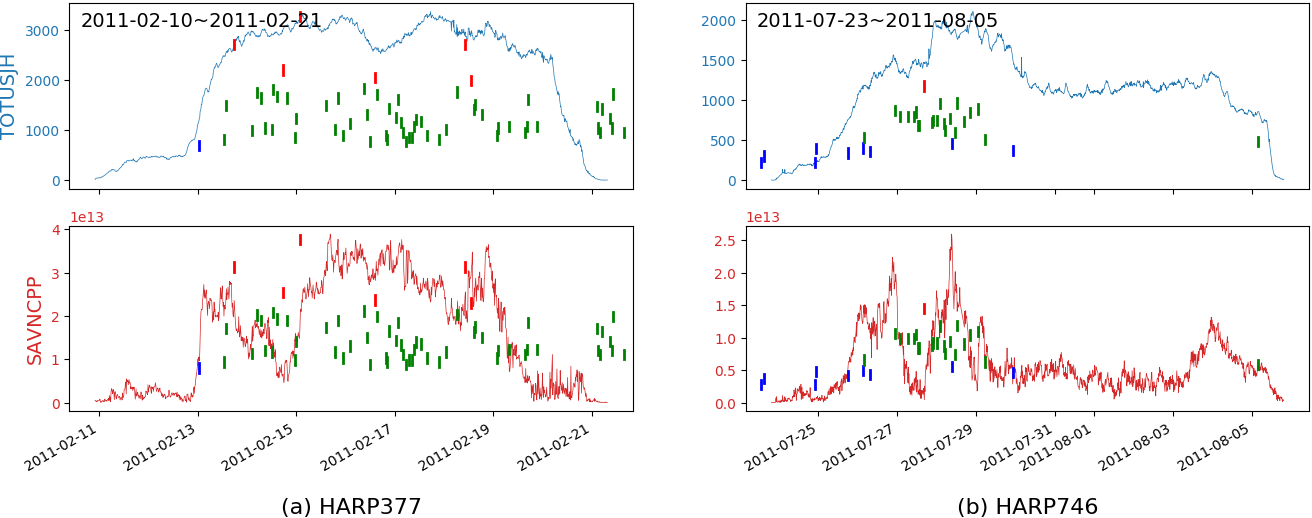}
    \caption{Examples of physical parameters derived from two HARPs, 377 and 746. The blue and red curves show the time variation of TOTUSJH and SAVNCPP quantities respectively.  Here, TOTUSJH stands for Total unsigned current helicity, and SAVNCPP stands for Sum of the modulus of the net current per polarity. Each small vertical line represents a recorded flare event. The height of the line is proportional to the $\log$ scale flare intensity, while red, green and blue represent M/X flare, C flare and B flare respectively.}
    \label{fig:exampleofar}%
\end{figure}

Fig.~\ref{fig:exampleofar} shows the flare history (B/C/M/X classes) for two HARPs (377 and 746) and time evolution of two important SHARP parameters, TOTUSJH and SAVNCPP, for a period of ten days (labeled on the x-axis).  {Specifically, TOTUSJH stands for Total unsigned current helicity, and SAVNCPP stands for Sum of the modulus of the net current per polarity.}  We can see that many incidences of C flares accompany a strong flare (of M/X class) and that the SHARP parameters evolve in continuous but locally stochastic ways during the energy buildup and release stages of strong flares. Therefore, it is important to consider the entire time series with flares of all classes, especially the highly prevalent C flares, when training machine learning models for flare prediction as opposed to only the time point where a weak (B) or strong (M/X) flare occurs as is done in~\citet{doi:10.1029/2019SW002214}. 

As found in the GOES data set, flares events occur sparsely, at irregular intervals, and at highly varying intensity levels, including long gaps between events, all of which present a unique challenge in the data analysis. We note that due to the fact that the amount of information contained in the observed data is limited, the inferential objective should be geared towards extracting the maximum amount of \textit{available} information and avoiding over-interpreting the data. Instead of seeking to model the flare intensity in continuous time for every time point, we model aggregated quantities instead, e.g. the maximum flare intensity within a fixed length time window (such as $\pm 12$ hours). In this way, we attach an intensity value to every data point that has a recorded flare in the neighboring $\pm 12$ hour time window. For the other time points, we define them as being ``quiet'' locally with an indicator function attached to it. We will explain the details of this data preparation process in Section~\ref{data_preprocess}. In our proposed prediction model, we are able to predict the maximum flare intensity level within a fixed length time window $T$ hours in the future, where $T$ can be specified to a desired value such as 12 or 24 hours, using the time series of SHARP parameters in the past. As a byproduct, we can classify the predicted events into strong or weak flares according to the flare level definitions. 

\section{Methodology}
\label{sec:methodology}
We provide a detailed description of the data pre-processing pipeline in Section~\ref{data_preprocess}. A mixed Long-Short Term Memory (LSTM) regression model (\citet{Hochreiter1997LongSM}) that can directly predict the solar flare intensity is introduced in Section~\ref{model_description}, including the model structure and a novel loss function. Section~\ref{class_models} covers three binary classification models based on the mixed LSTM regression model. They all try to distinguish the M and X flares from other flares (including or excluding the C flares) by making use of the predicted intensities given by the regression model.

\subsection{Data Preparation} \label{data_preprocess}

The machine learning models that we aim to train are prediction models, which require two sources of input data: the feature set (a.k.a. predictors) and the response variables. In this section, we give the details of the data sources and how we prepare the data for training and testing the machine learning models. 

For response variables, we use flare events recorded in the GOES data set ranging from 05/01/2010 to 06/20/2018 (MM/DD/YYYY). Within this time range there are a total of 12,012 recorded flares. See flare-event-only data set in Fig.~\ref{fig:dist} for the distribution of the flare events in GOES data set. Note that the theoretical distribution of the flare events should be a power law distribution. The reduced number of recorded flares in lower energy levels is because events are lost in the background and go undetected. Therefore, the observed distribution is different from the theoretical distribution and we are focused on the observed information in this paper.

For the source of data for features/predictors, we consider data from $860$ HMI Active Region Patches (HARPs). For the chosen time period (05/01/2010 to 06/20/2018), there are approximately 7000 HARPs, many occurring without flares. From these, in order to maintain the quality of the data, we down select the HARPs to a group of 860 based on the criteria (1) the longitude of the HARP should be within the range of $\pm 68^{\circ}$ from Sun central meridian,  {to avoid projection effects, see~\citet{Bobra_2015} and~\citet{doi:10.1029/2019SW002214}}; (2) the missing SHARP parameters should be fewer than 5 $\%$ of all in the HARP,  {to make sure that the missing data is not significantly large to cause any bias in model training}.  

For each HARP, there is a time series of vector magnetograms with 12-minute cadence. Here we consider the time series as a video with one frame every 12 minutes. We use the SHARP parameters, which are scalar variables derived from the full photospheric vector magnetic field. The SHARP parameters are calculated over the magnetogram of the each frame, see \citet{Bobra:2014a} for a detailed description of the calculations. Of all the SHARP parameters, we use USFLUX, MEANGAM, MEANGBT, MEANGBZ, MEANGBH, MEANJZD, TOTUSJZ, MEANALP, MEANJZH, TOTUSJH, ABSNJZH, SAVNCPP, MEANPOT, TOTPOT, MEANSHR, SHRGT45, SIZE, SIZE\_ACR, NACR and NPIX in our study  {(see the definitions of these parameters in Table~\ref{tab:SHARP_def})}. Therefore, each frame corresponds to one vector magnetogram and a $20\times 1$ SHARP vector. Each HARP corresponds to a data matrix with 20 columns and ``number of frames (vector magnetograms)'' rows. These data are provided by the Stanford Joint Science Operations Center (see \url{http://jsoc.stanford.edu}). 


\begin{table}[ht]
\centering
\begin{tabular}{ll}
\rowcolor[HTML]{C0C0C0} 
\textbf{Parameter} & \textbf{Description} \\
\rowcolor[HTML]{ECF4FF} 
TOTUSJH:           & Total unsigned current helicity \\
\rowcolor[HTML]{CBCEFB} 
TOTUSJZ:           & Total unsigned vertical current \\
\rowcolor[HTML]{ECF4FF} 
SAVNCPP:           & Sum of the modulus of the net current per polarity \\
\rowcolor[HTML]{CBCEFB} 
USFLUX:            & Total unsigned flux \\
\rowcolor[HTML]{ECF4FF} 
ABSNJZH:           & Absolute value of the net current helicity \\
\rowcolor[HTML]{CBCEFB} 
TOTPOT:            & Proxy for total photospheric magnetic free energy density \\
\rowcolor[HTML]{ECF4FF} 
SIZE ACR:          & \begin{tabular}[c]{@{}l@{}}De-projected area of active pixels ($B_{z}$ magnitude larger than\\ noise threshold) on image in micro-hemisphere (defined as\\ one millionth of half the surface of the Sun)\end{tabular} \\
\rowcolor[HTML]{CBCEFB} 
NACR:              & The number of strong LoS magnetic-field pixels in the patch \\
\rowcolor[HTML]{ECF4FF} 
MEANPOT:           & Proxy for mean photospheric excess magnetic energy density \\
\rowcolor[HTML]{CBCEFB} 
SIZE:              & Projected area of the image in micro-hemispheres \\
\rowcolor[HTML]{ECF4FF} 
MEANJZH:           & Current helicity (Bz contribution) \\
\rowcolor[HTML]{CBCEFB} 
SHRGT45:           & Fraction of area with shear $>45^{\circ}$ \\
\rowcolor[HTML]{ECF4FF} 
MEANSHR:           & Mean shear angle  \\
\rowcolor[HTML]{CBCEFB} 
MEANJZD:           & Vertical current density \\
\rowcolor[HTML]{ECF4FF} 
MEANALP:           & Characteristic twist parameter, $\alpha$ \\
\rowcolor[HTML]{CBCEFB} 
MEANGBT:           & Horizontal gradient of total field \\
\rowcolor[HTML]{ECF4FF} 
MEANGAM:           & Mean angle of field from radial \\
\rowcolor[HTML]{CBCEFB} 
MEANGBZ:           & Horizontal gradient of vertical field \\
\rowcolor[HTML]{ECF4FF} 
MEANGBH:           & Horizontal gradient of horizontal field \\
\rowcolor[HTML]{CBCEFB} 
NPIX:              & Number of pixels within the patch                     \end{tabular}
\caption{List of SHARP parameters and brief descriptions. }
\label{tab:SHARP_def}
\end{table}
\subsubsection{Response Variable} \label{response_variable} 
Since some of the flares recorded in the GOES data set happened in HARPs that are not recorded in the filtered JSOC data, we consider 10,349 out of the total 12,012 flares recorded in the GOES data set during the time range indicated on Table~\ref{tab:flares_num_years}.  {Moreover, the flares recorded in the GOES data set are listed by NOAA active region numbers while the corresponding photospheric magnetic field is identified with HARP patches, which use different criteria to identify and group the strong field regions. Consequently, there is the potential issue of a single HARP corresponding to multiple active regions; in fact, roughly 20\% of SHARP patches include components from multiple active regions. This problem has been acknowledged in~\citet{doi:10.1029/2019SW002214} and more details can be found therein. In this paper, we do not address this potential problem caused by the data but focus on the methods for modeling. We speculate that this potential problem of mismatch of SHARP and GOES data may or may not result in biases for prediction models, while might incur loss of statistical efficiency due to the extra noise brought in. }

In order to make maximum use of the data, we consider not only the class of each flare, but also the exact value of the flare intensity whkich is defined as the peak flux in watts per square metre (W/$m^2$) of soft X-rays with wavelengths 100 to 800 picometres. Moreover, since the flare intensity spans orders of magnitude, we take the $\log_{10}$ transform (see Table~\ref{take_log}) in order to better handle the extreme values, X and M flares. All flare intensities mentioned later are $\log_{10}$ scale intensities if not further specified.

\begin{table}[ht]
\centering
\begin{tabular}{cccclllllll}
\rowcolor[HTML]{C0C0C0} 
\textbf{Class/Year} & 2010 & 2011 & 2012 & 2013 & 2014 & 2015 & 2016 & 2017 & 2018 & Total \\
\rowcolor[HTML]{ECF4FF} 
X                   & 0    & 8    & 5    & 12   & 15   & 2    & 0    & 4    & 0    & 46    \\
\rowcolor[HTML]{CBCEFB} 
M                   & 8    & 84   & 110  & 90   & 169  & 128  & 7    & 37   & 0    & 633   \\
\rowcolor[HTML]{ECF4FF} 
C                   & 64   & 788  & 906  & 1105 & 1231 & 1194 & 244  & 225  & 11   & 5768  \\
\rowcolor[HTML]{CBCEFB} 
B                   & 512  & 519  & 398  & 418  & 94   & 428  & 722  & 606  & 205  & 3902 
\end{tabular}
\caption{The number of X/M/C/B flares recorded in each year in the GOES data set during the time range 05/01/2010-06/20/2018. }
\label{tab:flares_num_years}
\end{table}

\begin{table}[ht]
\centering
\begin{tabular}{llll}
\rowcolor[HTML]{C0C0C0} 
Flare class & Peak flux range ($\text{W}/m^2$) & \cellcolor[HTML]{FFFFFF}      & $\log_{10}$ intensity  \\
\rowcolor[HTML]{CBCEFB} 
X & $\geq10^{-4}$        &\cellcolor[HTML]{FFFFFF} 
&       $ \geq -4$ \\
\rowcolor[HTML]{ECF4FF} 
M & $10^{-5}\sim10^{-4}$ & \cellcolor[HTML]{FFFFFF}  & $-5 \sim -4$ \\
\rowcolor[HTML]{CBCEFB} 
C & $10^{-6}\sim10^{-5}$ &  \cellcolor[HTML]{FFFFFF}                                                  & $-6 \sim -5$ \\
\rowcolor[HTML]{ECF4FF} 
B & $10^{-7}\sim10^{-6}$ &\cellcolor[HTML]{FFFFFF} 
\multirow{-4}{*}{$\xrightarrow{\text{$\log_{10}$}}$} & $-7 \sim -6$      
\end{tabular}
\caption{Transformation from flares class to continuous intensity values we adopt.}
\label{take_log}
\end{table}

After performing the data processing as described above, there are over 10,000 flares identified from a time history of X-ray intensity levels. However, considering only the peak intensity level recorded at a given time point as in~\citet{doi:10.1029/2019SW002214}, there are some limitations, stated below.

\begin{enumerate}
    \item Most of the M and X flare events are accompanied by much more frequent C flares. If we simply assign the response variable based on flares' peak times, two flares happening adjacent to each other with totally different intensities can have a large amount of overlapping training data (time series). Two observations with similar training data but quite different response variables would confuse the model.
    \item Even though there are over 10,000 flare records in GOES data set, they are not all in the recorded range of the 860 HARP videos.  {Also, the number of the strong flares which we care the most are limited (see Table~\ref{tab:flares_num_years}).} Besides, some of the HARP videos are not suitable for use in training machine learning models due to large amounts of missing entries in the SHARP parameters. Therefore, the effective number of flare events that we can use for training/testing the machine learning model is not as large as expected. 
    \item The recorded flares only occupy a very small fraction of the time series of observations, i.e. the SHARP parameters. Those time points without a recorded flare might be an unrecorded weak flare near a stronger one, or most likely a ``flare-free'' time point.   {Considering these time points as contrasts to the time points with flares can help the model better distinguish the strong flares from the others. } {Therefore, discarding this piece of information would impair the performance of the prediction model.}
\end{enumerate}

\begin{figure}[htb]
\centering
\includegraphics[width=0.7\textwidth]{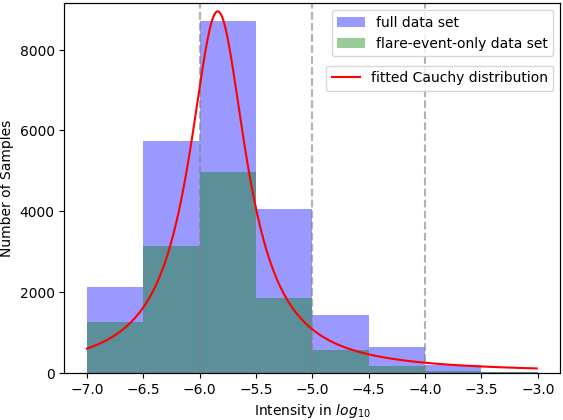}
\caption{The distribution of non-quiet samples' flare intensities ($I$) in flare-event-only data set and full data set, where flare-event-only data set only takes flare intensities recorded on GOES data set as response variables. The definition of full data set can be seen in Section \ref{response_variable}. Red line is the fitted Cauchy distribution with location parameter $x_0$ = -5.84 and scale parameter $\gamma$=0.31.}
\label{fig:dist}
\end{figure}

Therefore, in order to overcome these drawbacks, we propose the following definition of response variables in our prediction model: for each frame, we define its real-time intensity as the maximum flare intensity that happened within a 24-hour time window (12 hours before and 12 hours after). In other words, instead of focusing on each recorded flare in GOES data set, we only care about the largest flare that happened in each frame's 24-hour time window. By applying this new mechanism, we can assign each frame a response variable. Correspondingly, the new data set is called ``full data set'' (see the distribution of the flares in the constructed full data set as compared to the flare-event-only data set in Fig.\ref{fig:dist}). As a result, the non-quiet sample size of the full data set is over two times larger as compared to the flare-event-only data set, 22,928 as opposed to 10,349. Plus, the response variables of those C flares happening next to strong flares (M or X) are redefined as high intensities which is certainly more reasonable for model training. Most importantly, this mechanism more accurately portray {s} the processes of solar activities: instead of being single-time-point incidences, they are processes of extended time evolution.

A natural question is how to deal with the frames where there is no flare recorded in the 24-hour time window. We define one more binary response variable to denote the ``flaring'' or ``non-flaring'' of the 24-hour time window -- 1 means there is at least one flare (M/X/C/B-class) recorded in the GOES data set within the 24-hour window while 0 means no flares recorded in the GOES data set within the window.

\begin{figure}[htb]
    \centering
    \includegraphics[width=0.7\textwidth]{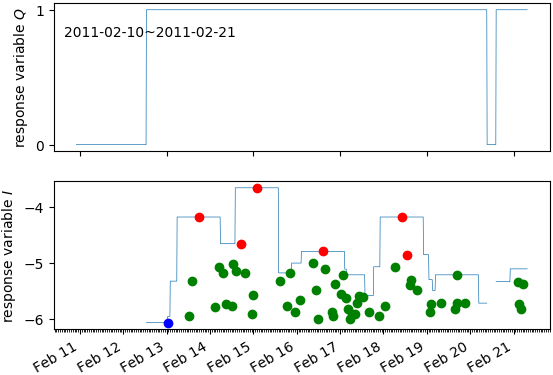}
    \caption{An example of how we define response variables based on the recorded flares that happened with HARP 377. The lower panel is the value of $I$ given all flares while the upper panel is the value of $Q$. Still, red, green and blue points represent M/X, C and B flares respectively. Notice that there  {are} missing values of $I$. The missing part is defined as the quiet region where correspondingly $Q$s take  {a} value of 0. }
    \label{fig:define_response_exm}
\end{figure}

\begin{table}[htb]
\centering
\begin{tabular}{lc}
\rowcolor[HTML]{C0C0C0} 
Label & Response Variable ($[Q,I]$) \\
\rowcolor[HTML]{CBCEFB} 
M1.5  & [1, -4.824]              \\
\rowcolor[HTML]{ECF4FF} 
X1.6  & [1, -3.796]           \\
\rowcolor[HTML]{CBCEFB} 
C7.2  & [1, -5.143]     \\
\rowcolor[HTML]{ECF4FF} 
Quiet     & [0, $N/A$]        
\end{tabular}
\caption{Examples of the how we define response variables given the flare labels. Quiet stands for one quiet sample. See Section~\ref{response_variable} for details.}
\label{example_trans}
\end{table}

To recap, for each frame, we assign it a 2-dimensional response variable, the first dimension $Q$ corresponds to the ``local quietness'' or ``local non-quietness'' (Boolean, 1 for having a flare event within the 24-hour window and 0 for not having a flare event within the 24-hour window) while the second dimension $I$ stands for its real-time intensity on the $\log_{10}$ scale (continuous). Specifically, if a sample has $Q=0$, then we annotate the second dimension of its response variable as $N/A$ (see Table~\ref{example_trans}). An example of how we define the response variable $[Q,I]$ for HARP 377 is shown in Fig.~\ref{fig:define_response_exm}.


\begin{figure}[htb]
\centering
\includegraphics[scale=0.4]{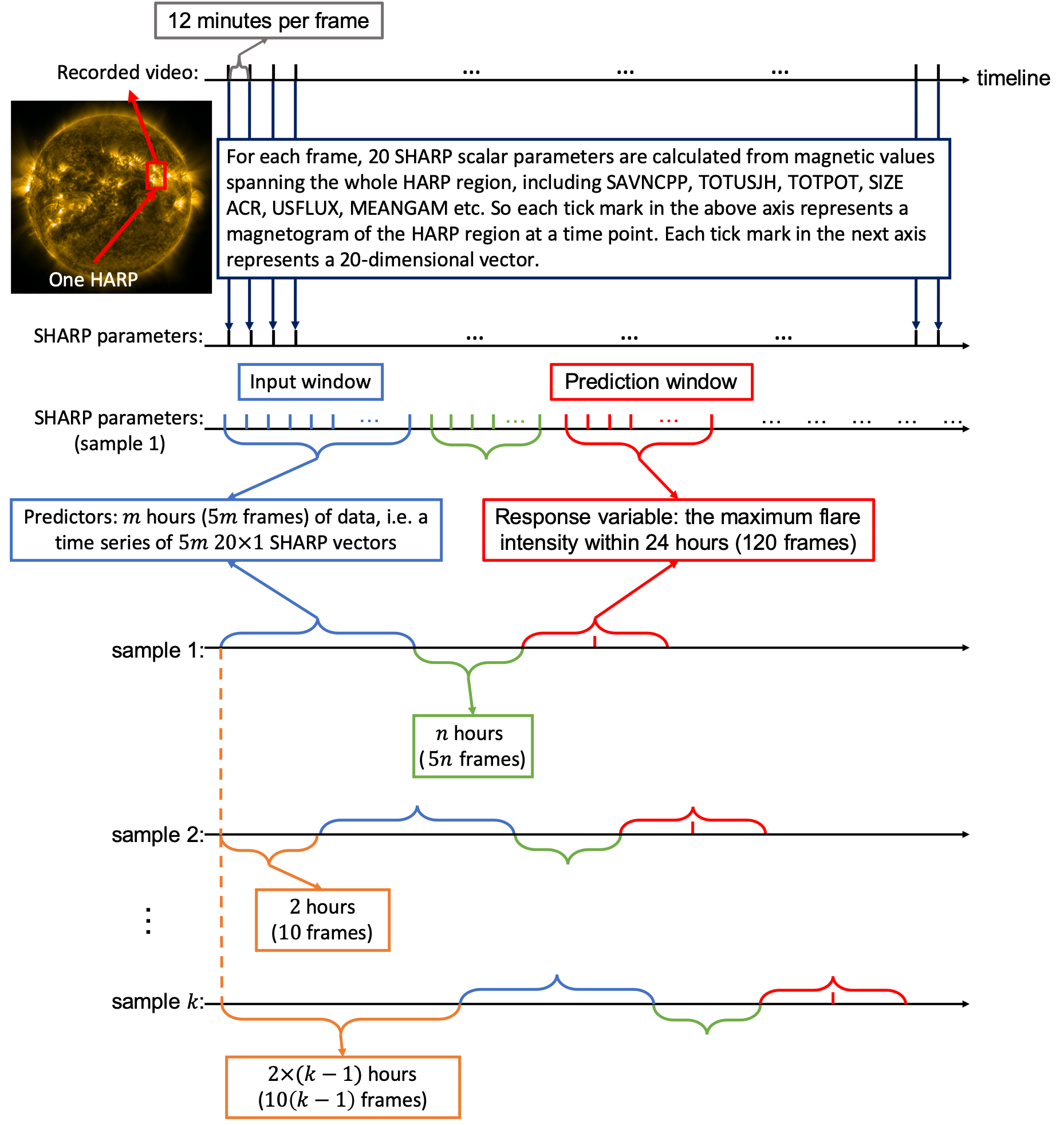}
\caption{A diagram of how we prepare samples for training the algorithm (See Section~\ref{pipeline}). For each HARP, there is a ``video'' containing a time series of magnetograms. For each frame, 20 SHARP parameters are calculated from the magnetic field components over the whole HARP. Therefore, we can obtain a data matrix for each HARP with 20 columns and  ``the number of frames (magnetograms)'' rows. Data in blue braces are the predictors. Green braces denote the prediction intervals and the response variables are decided based on the maximum flare intensities recorded in red braces. Samples are taken every 10 frames.}
\label{get_sample}
\end{figure}


\subsubsection{Input Data Pre-processing Pipeline} \label{pipeline} 

A detailed diagram of how we prepare the raw data for machine learning is shown in Fig.~\ref{get_sample}. We briefly describe it here. Suppose we aim to train a model that uses $m$ hours of SHARP parameters to predict the maximum flare intensity in the 24-hour window beginning at $n$ hours after. Since the time cadence of our data is 12 minutes, there are 5 observed frames (magnetograms) at each hour. Each video needs to contain $5\times \left(m+n+24\right)$ consecutive frames to have at least one sample available. We take samples every 2 hours (10 frames), a reasonable step size which is neither too long to capture the detailed behaviors of the HARP nor so short that it causes oversampling of the time series. We take HARP 394 as an example. There are 1,334 frames in total. The training samples include frame 0 $\sim$ frame $5m-1$, frame 10 $\sim$ frame $5m+9$, ... , frame $10k$ $\sim$ frame $5m+10k-1$, ... Correspondingly, the response variables include the maximum flare intensities recorded within frame $5(m+n)$ $\sim$ frame $5(m+n+24)-1$, frame $5(m+n)+10$ $\sim$ frame $5(m+n+24)+9$, ... , frame $5(m+n)+10k$ $\sim$ frame $5(m+n+24)+10k-1$, ..., where $k=0,1,2...$ and $5(m+n+24)+10k-1<1334$.

\label{train_test_split}

We split the training and testing data by years in order to avoid information leaking. Since all the recorded data ranges from 2010 to 2018, we have that roughly 63\% of flares happened before 2015 (6,536 out of 10,349). We note that the corresponding sample size as obtained by the data preparation described above has a similar flare rate. Each HARP only has one video, so no HARP is divided in both the training and testing set. In this study, we split all flares that happened before 01/01/2015 into the training set and the rest into the testing set. After splitting the data into training/testing samples, we normalize all the data by subtracting the mean and dividing by the standard deviation computed from the training data \citep[Chapter~7.10]{hastie2009elements}. No information from the testing data is used in the normalization step.


Some of the HARPs have missing frames, which result in the time interval between two adjacent frames being longer than 12 minutes. In this case, we set up a tolerance threshold: if the number of missing frames in total for one sample input is less or equal to 10, we apply hot deck imputation \citep{andridge_little_2010} to fill the missing values. However, if there are more than 10 frames missing, we drop the sample.


\subsection{Model Description} \label{model_description}
We adopt a mixed LSTM (\citet{Hochreiter1997LongSM}) regression model to portray the relationship between SHARP parameters and flares, with a novel loss function to measure the differences between predicted results and the 2-dimensional response variables defined in Section~\ref{response_variable}. The LSTM model predicts outcomes using trained non-linear transformations of input parameters and has been applied to classification of time-series data \citep[Chapter~10]{goodfellow2016deep}. It should be noted that in \citet{doi:10.1029/2019SW002214}, the LSTM is only used for binary classifications whereas in this paper, the LSTM is used for both regression and classification.  {We call the proposed model a mixed LSTM regression model in that it is an LSTM model combining regression and classification tasks.}

\subsubsection{Model Structure} \label{model_structure_def}
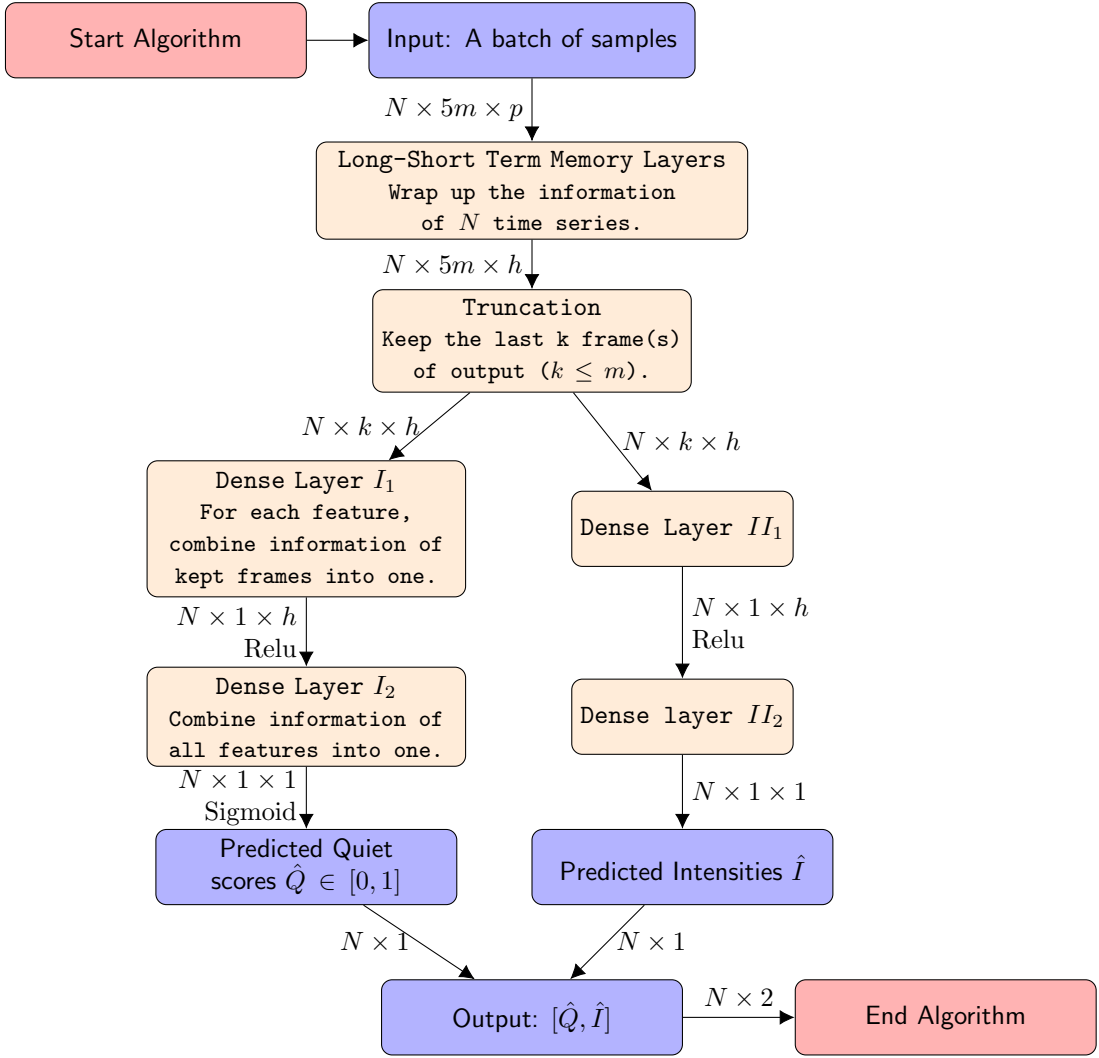
\begin{figure}[ht]
\centering
\begin{tikzpicture}[node distance=2cm]
\node (start)[startstop]{Start Algorithm};
\node (in1) [io, right of=start, text width=4.1cm, xshift = 3cm] {Input: A batch of samples};
\node (pro1) [process, below of=in1, text width=5.5cm] {Long-Short Term Memory Layers \\ \small Wrap up the information of $N$ time series.};
\node (pro11)[process, below of=pro1, text width=4cm]{Truncation \\ \small Keep the last k frame(s) of output ($k\leq m$).};
\node (pro2) [process, below of=pro11, xshift=-3cm, yshift=-0.5cm, text width=4cm]{Dense Layer $I_1$\\ \small For each feature, combine information of kept frames into one.};
\node (pro3) [process, right of=pro2, xshift=3cm]{Dense Layer $II_1$};
\node (pro5) [process, below of=pro2, yshift = -0.5cm, text width=4cm]{Dense Layer $I_2$ \\ \small Combine information of all features into one.};
\node (pro6) [process, right of=pro5, xshift=3cm]{Dense layer $II_2$};
\node (out1) [io, below of=pro5, text width = 3.5cm]{Predicted Quiet scores $\hat{Q}\in[0,1]$};
\node (out2) [io, right of=out1, xshift=3cm, text width = 3.5cm]{Predicted Intensities $\hat{I}$};
\node (out3) [io, below of=out1, xshift=3cm]{Output: $[\hat{Q},\hat{I}]$};
\node (end)[startstop, right of=out3, xshift=3.5cm]{End Algorithm};

\draw [->] (start) -- (in1);
\draw [->] (in1) -- node[anchor=east]{$N \times 5m \times p$} (pro1);
\draw [->] (pro1) -- node[anchor=east]{$N \times 5m \times h$} (pro11);
\draw [->] (pro11) -- node[anchor=east]{$N \times k \times h$}(pro2);
\draw [->] (pro11) -- node[anchor=west]{$N \times k \times h$}(pro3);
\draw [->] (pro2) -- node[anchor=east, text width=1.8cm,align=right] {$N \times 1 \times h$ \\Relu}(pro5);
\draw [->] (pro3) -- node[anchor=west, text width=1.8cm, align=left] {$N \times 1 \times h$ \\Relu}(pro6);
\draw [->] (pro5) -- node[anchor=east, text width=1.8cm, align=right] {$N \times 1 \times 1$ \\Sigmoid}(out1);
\draw [->] (pro6) -- node[anchor=west, text width=1.8cm,align=left] {$N \times 1 \times 1$}(out2);
\draw [->] (out1) -- node[anchor=east]{$N \times 1$}(out3);
\draw [->] (out2) -- node[anchor=west]{$N \times 1$}(out3);
\draw [->] (out3) -- node[anchor=south]{$N \times 2$}(end);
\end{tikzpicture}
\caption{The flowchart of the LSTM regression model, discussed in Section~\ref{model_structure_def}. In the figure, $N$ is the number of samples in one batch, $5m$ is the number of frames for each sample (see Fig.\ref{get_sample} for details), and $p$ is the number of features we take into consideration. $h$ is the dimensionality of the LSTM layers and the output space and $k$ is the number of frame(s) we keep after going through the LSTM layers. }
\label{lstm_alg}
\end{figure}

The flowchart of the model is shown in Fig.~\ref{lstm_alg}. For each sample, the input/predictor is $5m$ sets of SHARP parameters (see Fig.\ref{get_sample}), a $1 \times 5m \times p$ tensor. Again, $m$ is the number of hours of data we use for prediction before current time point and $n$ is number of hours from 24-hour window's left bound to now. $m$ takes value from 6, 12, 24 and 48, which are a series of data lengths typically considered for training prediction models for solar flares; $n$ takes values from 0, 6, 12, 24; and $p$ takes the value of 20, since we consider 20 SHARP parameters. The output/response is a $2 \times 1$ vector, including the predicted quiet score, $\hat{Q}$ and predicted intensity, $\hat{I}$ (see Table~\ref{example_trans}).

As shown in Fig.~\ref{lstm_alg}, the model starts with LSTM layers. We introduce dropout layers \citep{JMLR:v15:srivastava14a} between adjacent LSTM layers with dropout ratio = 0.3. The number of LSTM layers = 4, the dimensionality $h$ of the LSTM layers and the output space is $30$, and the sample size $N$ in one batch is set to be 40. Take  {a} model with $m=24$ and $n=6$ as an example. We have 38,906 samples available in training set (see Section~\ref{pipeline}). For each epoch, we randomly assign them to $41869/40 \approx 973$ batches. Therefore, the input is one batch out of 973, a $40\times120\times20$ tensor. After the LSTM layers, the output is a $40\times120\times30$ tensor, given $h = 30$. Then, it goes through the truncation procedure, during which the tensor becomes $40\times k \times30$, typically $k<< 120$. Considering that LSTM is a sequential model for time series \citep[Chapter~10]{goodfellow2016deep}, the choice of $k=5m=120$ corresponds to the sequence prediction model that explicitly adopts all these 120 input frames. However, our main goal is to capture the behavior of the $5n$ subsequent HARP frames. Therefore, the output from the latter few frames ($k$ frames) suffice for making the desired predictions. Specifically, $k$ takes the value of 1 in our models. Nevertheless, we have tried taking more than one ($k = 2, 5, 10...$) frames' output into the next layer and did not obtain a significantly better result.

After the LSTM and truncation layers, we feed it to two separate sub-models for $Q$ and $I$'s training respectively, each of which contains two dense layers. The first dense layer serves the purpose of reducing the second dimension of the tensor to 1, while the second condenses the third dimension to 1. Intuitively, the first dense layer works to combine all the information in all $k$ frames to 1 frame for each feature and the second combines information of all $p$ features into 1 super-feature. A \texttt{Relu} function is added between two dense layers to break the linearity. Since we take $k=1$ in our models, the Dense Layer $I_1$ and $II_1$ shown in Fig.~\ref{lstm_alg} are deprecated, leaving only \texttt{Relu} functions. The only difference between these two sub-models is that we further add a \texttt{Sigmoid} function at the end of the Q-training model in order to keep its value, interpreted as the probability of being unquiet, between $[0,1]$. Though $Q$ and $I$ go through two separate pipelines, they are not independent during the training. We introduce the loss function in Section~\ref{loss_function_des} that enables us to consider $Q$ and $I$ jointly in the training.

\begin{figure}[ht]
    \centering
    \includegraphics[width=0.7\textwidth]{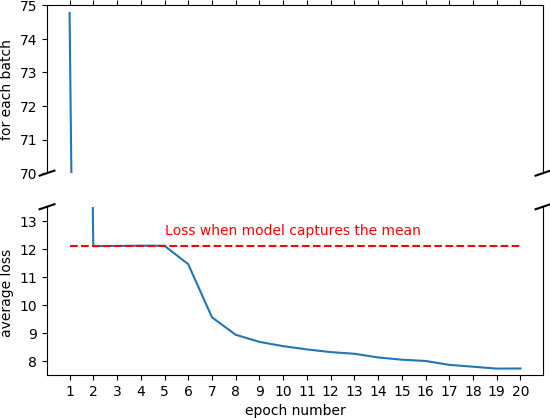}
    \caption{An example showing the convergence behavior of the mixed LSTM regression model. The x-axis labels the epoch number and the y-axis stands for the average loss across batches. From epoch 1 to epoch 2, the average loss for each batch drops approximately from 75 to 12. In order to also visualize clearly the super-linear change starting at epoch 5 in one figure, we cut the intermediate part of the loss change between epochs 1 and 2.}
    \label{fig:loss_change}
\end{figure}

We set the epoch number to be 20. Each model takes 5-7 epochs, which costs 5 to 10 minutes, to converge; and around 20 minutes to finish all the 20 epochs (on a 2.3GHz, i5, 16GB machine that we use). Typically, during the first 1-3 epochs, the model learns the means of all response variables and assigns the predicted intensities as the sample mean. Then, it takes a few epochs for the model to optimize over the parameters. And in the next 1-3 epochs, the loss converges super-linearly. Fig.~\ref{fig:loss_change} gives a typical example of the variation of the loss function in the training process. We will give a detailed definition of the loss function in Section~\ref{loss_function_des}.

 {
Specifically, we here reemphasize several strategies implemented to avoid overfitting issues. First, the dropout layers with dropout ratio equal to $0.3$ are set between adjacent LSTM layers. Those dropout layers randomly rule out 30\% of the neurons from the preceding LSTM layers which have been proven to be an efficient way to avoid overfitting~\citep{srivastava2014dropout}. Second, we apply early stopping with back propagation strategy~\citep{doan2004generalization} by setting the epoch number to 20. Last and most importantly, the sample size is over 60,000 -- 37,784 quiet samples plus 22,928 non-quiet samples -- after the pre-processing pipeline (Section~\ref{pipeline}), which is enough for model to learn the behavior of solar flares comprehensively.}

\subsubsection{Loss Function} \label{loss_function_des}

In our mixed LSTM regression model, the response variables contain both Boolean and continuous values. Therefore, we need to adopt a special mixed approach to jointly evaluate the loss. In addition, for those samples with $Q=0$, there are no exact values of intensity recorded. We assign $N/A$ to those ``missing'' intensity values. The desired loss function should avoid the usage of $I$ for those samples with intensity values missing. We use binary cross-entropy loss in terms of $\hat{Q}$, which takes values between $0$ and $1$; and the squared error loss for $\hat{I}$ (\citet{Janocha2017}), which takes values in $\mathbbm{R}$; see Table~\ref{tab:loss} for examples. Furthermore, we define three tuning parameters to flexibly deal with the overabundance of the quiet samples and the non-comparability between the loss for quiet score and that for (logarithm) intensity values.

\begin{table}[ht]
\centering
\begin{tabular}{lcc}
\rowcolor[HTML]{C0C0C0} 
Loss & Quiet Sample & Non-quiet sample \\
\rowcolor[HTML]{CBCEFB} 
$Q$    & $-\log(1-\hat{Q}$)     & $-\log(\hat{Q})$            \\
\rowcolor[HTML]{ECF4FF} 
$I$    & $N/A$                & $(I-\hat{I})^2$          
\end{tabular}
\caption{We use binary cross-entropy loss in terms of $\hat{Q}$ and $L_2$ loss for $\hat{I}$}
\label{tab:loss}
\end{table}

More precisely, the loss function for each batch is defined as 
\begin{align*}
    \mathcal{L} = &\sum_{\rm batch\ samples}^{N}  
    \left[ \begin{array}{cc} r & 1 \end{array} \right]
    \begin{bmatrix}
        -\log(1-\hat{Q})&-\log(\hat{Q})\\
        0 &(I-\hat{I})^2\\
    \end{bmatrix} 
    \left[ \begin{array}{c} \mathbbm{1}(Q = 0)w_1 \\ \mathbbm{1}(Q \neq 0)w_2(I) \end{array} \right] 
    \\
    = &\sum_{\rm batch\ samples}^{N} \left[-\mathbbm{1}(Q = 0) w_1 r\log (1-\hat{Q}) + \mathbbm{1}(Q \neq 0) w_2(I) (-r \log \hat{Q} + (I - \hat{I})^2)\right],  
\end{align*}
where $Q$ only takes values in the binary set $\{0, 1\}$, $I\in[-7,-3]$ are observed log-intensity values, $\hat{Q}\in [0, 1]$ and $\hat{I}\in\mathbbm{R}$ are fitted values, $\mathbbm{1}(Q = 0)$ is the indicator function for $Q=0$, and $N$ is the sample size of each batch. We take $N=40$ in all our models (see Section~\ref{model_structure_def}). The tuning parameters $w_1$, $w_2(\cdot)$ and $r$ are adopted to calibrate the weight of each component in the loss function. Specifically, $w_1$ is the weight for loss generated by quiet samples, while $w_2(.)$ is a function set for non-quiet samples returning weights given specific intensity, and $r$ is the weight for the loss generated by the $Q$ dimension. Note that for the loss function, only the relative values of $w_1, w_2(\cdot)$ and $r$ matter -- a loss function can be defined up to a positive constant. Next we explain the different components in the design of this loss function.


For the loss generated by the $Q$ dimension, since $Q\in\{0,1\}$ and $I\in[-7,-3]$, the scale of $Q$'s loss is incomparable to $I$'s loss. We multiply the $Q$ dimension's loss by a scale parameter $r$ for all samples in order to balance the losses of $Q$ and $I$. In terms of loss of quiet samples, there are significantly more of them {, 37,784,} than non-quiet samples (flare events) {, 22,928.} We note that our main focus is on those non-quiet samples when predicting local maximum flare intensities. Therefore, we multiply the loss of the quiet samples with weight $w_1(<1)$ in order to attenuate the impact caused by the overabundance of quiet samples when training our prediction models. The values of $r$ and $w_1$ are both tuned by the cross-validation (\citet[Chapter~7.10]{hastie2009elements}) {.}  {Specifically, we consider $r$ taking values in set $\{1,2,5,10,15\}$ and $w_1$ taking values in set $\{0.1,0.2,0.5,1\}$. We randomly divide the training data set into 10 folds. For each possible pair of $r$ and $w_1$, we train the model 10 times with 9 folds as the training set and the remaining fold as the testing set. Finally we take the parameter values $r=5, w_1=0.2$, which results in the lowest average loss. }

Now we consider the loss associated with the non-quiet samples (flare events). As we can see in Fig.~\ref{fig:dist}, C flares dominate the data set while the samples for B and M/X flares are comparatively more limited. We adopt the squared error loss for the prediction of flare intensities. If we simply weight all the input samples equally, under the square loss setting, the consequence is that the predicted results will tend to cluster at the central part (around -6 to -5.5 for logarithm intensity, corresponding to C flares), which are the 30\% and 70\% quantiles of the response variables respectively, instead of the [-7, -3] intensity range. 
This is inconsistent with our original intention that M/X flares shall stand out from other flares as much as possible in the model. Thus we add $w_2(\cdot)$ (see Eq.~\eqref{w_2}) which serves to balance the weights of samples from different classes, which down-weights the prevalent C flares essentially. We define the weight for the flare with intensity level $I$ as
\begin{linenomath*}
\begin{equation}
w_2(I) = |I - \mu| \times \text{constant}.
\label{w_2}
\end{equation}
Next we explain our rationale for choosing this particular set of weights. We fit the empirical distribution of the logarithm of the flare intensity of the full data set to a Cauchy distribution, which is a heavy-tailed distribution, with location parameter $\mu$= -5.84 and scale parameter $\gamma$ = 0.31. The fitted curve is shown in Fig.~\ref{fig:dist}. The weight is set to be the $L_1$ distance from $\mu$ multiplied by a constant specified based on the proportion of the quiet samples. By doing so, we maintain the balance of samples of M/X, C and B classes. Eq.~\eqref{w_3} gives the detailed probability mass corresponding to each flare class under the weighting scheme given by Eq.~\eqref{w_2}:
\begin{equation}
\left\{
\begin{array}{lr}
{\rm B\ flares:}\quad\int_{-7}^{-6} |x - \mu|\cdot f(x) dx  = 0.121 \\
{\rm C\ flares:}\quad \int_{-6}^{-5} |x - \mu|\cdot f(x) dx = 0.116 \\
{\rm M/X\ flares:}\quad\int_{-5}^{-3} |x - \mu|\cdot f(x) dx = 0.114
\end{array}
\right.,
\label{w_3}
\end{equation}
\end{linenomath*}
where a Cauchy distribution with location parameter $\mu$ and scale parameter $\gamma$ has probability density function denoted by $f(x) =\left[\pi \gamma (1+\left(\frac{(x-\mu)}{\gamma})^2\right)\right]^{-1}$. 

With this strategy, we can combine the quiet and non-quiet samples in one model and train them simultaneously. Again, the loss function $\mathcal{L}$ is defined over each batch with $N$ samples therein. Therefore, we can obtain the ``number of batch'' of losses for each epoch. The loss we evaluate and visualize in Fig.~\ref{fig:loss_change}  is the average loss of all batches over each epoch. The results calculated based on the loss function $\mathcal{L}$ are shown in Section~\ref{reg_result_sec}.



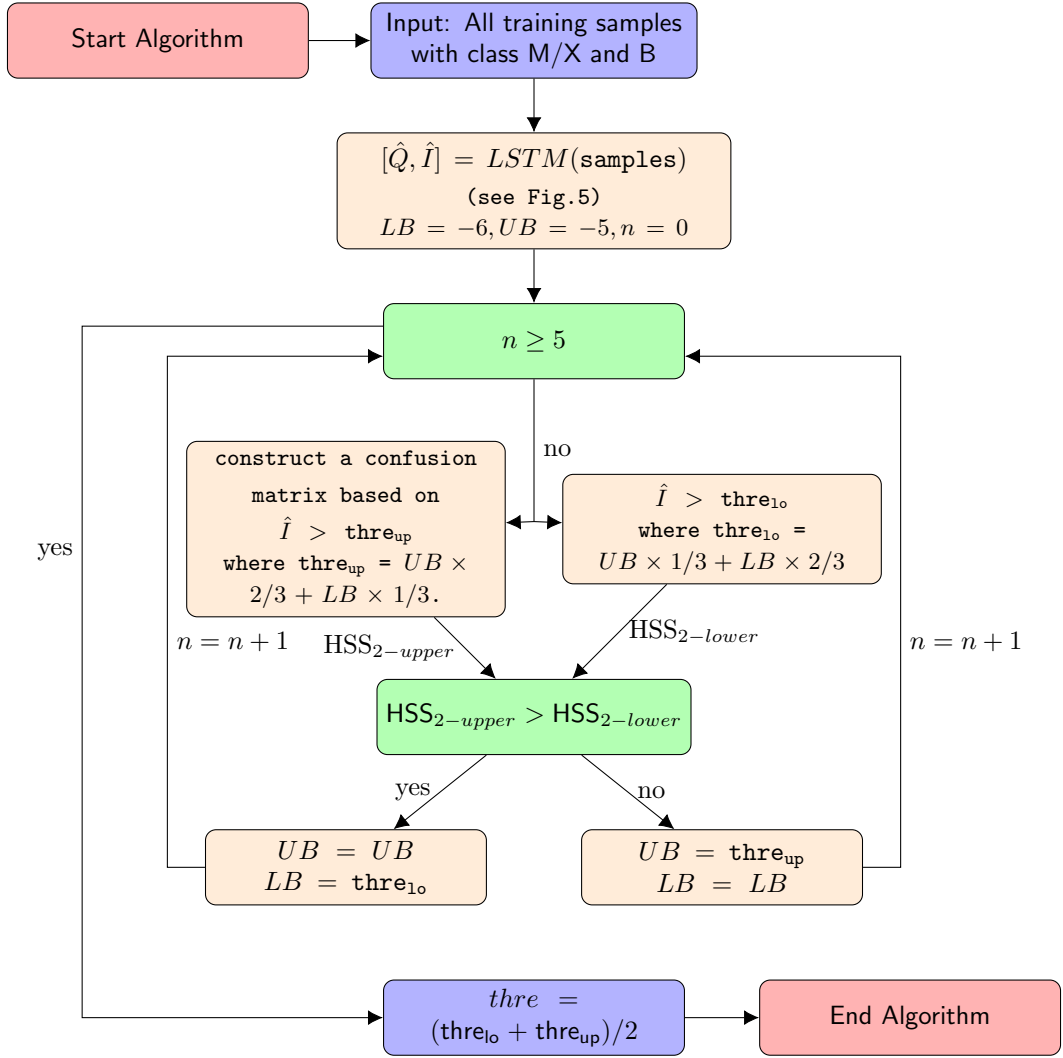
\begin{figure}[ht]
\centering
\begin{tikzpicture}[node distance=2cm]
\node (start)[startstop]{Start Algorithm};
\node (in1) [io, right of=start, text width=4.1cm, xshift = 3cm] {Input: All training samples with class M/X and B};
\node (pro1) [process, below of=in1, text width=5cm] {$[\hat{Q},\hat{I}] = LSTM(\text{samples})$  \\ \small (see Fig.\ref{lstm_alg}) \\$LB=-6, UB=-5, n=0$};
\node (dec0)[decision, below of=pro1]{$n\geq5$};
\coordinate [below of=dec0, yshift=-0.4cm] (d1);
\node (dec1) [process, below of=dec0, xshift=-2.5cm, text width=4cm, yshift = -0.5cm]{\small construct a confusion matrix based on \\
$\hat{I} > \text{thre}_\text{up}$ \\ where
$\text{thre}_\text{up}$ = $UB\times 2/3 + LB\times 1/3$.};
\node (dec2) [process, right of=dec1, xshift=3cm, text width=4cm]{\small 
$\hat{I} > \text{thre}_\text{lo}$ \\  where
$\text{thre}_\text{lo}$ = $UB\times 1/3 + LB\times 2/3$};
\node (dec3) [decision, below of=dec1, xshift=2.5cm, yshift=-0.5cm]{$\text{HSS}_{2-upper} >\text{HSS}_{2-lower}$};
\node (pro2) [process, below of=dec1, text width = 3.5cm, yshift = -2.5cm]{$UB = UB$ \\ $LB = \text{thre}_\text{lo}$};
\node (pro3) [process, right of=pro2, xshift=3cm, text width = 3.5cm]{$UB = \text{thre}_\text{up}$ \\ $LB = LB$};
\node (pro4) [io, below of=pro2, xshift=2.5cm, text width = 3.5cm]{$thre = (\text{thre}_\text{lo} + \text{thre}_\text{up})/2$};
\node (end)[startstop, right of=pro4, xshift=3cm]{End Algorithm};

\draw [->] (start) -- (in1);
\draw [->] (in1) -- (pro1);
\draw [->] (pro1) -- (dec0);
\path [] (dec0) edge node[anchor=west] {no}(d1);
\draw [->] (d1) -- (dec1);
\draw [->] (d1) -- (dec2);
\draw [->] (dec1) -- node[anchor=east]{$\text{HSS}_{2-upper}$}(dec3);
\draw [->] (dec2) -- node[anchor=west]{$\text{HSS}_{2-lower}$}(dec3);
\draw [->] (dec3) -- node[anchor=east] {yes}(pro2);
\draw [->] (dec3) -- node[anchor=west] {no}(pro3);
\draw [->] (pro4) -- (end);
    
\draw[->]  
    (pro2.west) 
    -- + (-0.5, 0cm)
    -- + (-0.5, 3cm)
        node[anchor=west] {$n=n+1$}
    |- ([yshift=-0.2cm] dec0.west);
\draw[->]  
    (pro3.east) 
    -- + (0.5, 0)
    -- + (0.5, 3)
        node[anchor=west] {$n=n+1$}
    |- ([yshift=-0.2cm] dec0.east);
\draw[->]  
    ([yshift=0.2cm] dec0.west)
    -- + (-4, 0)
    -- + (-4, -3)
            node[anchor=east] {yes}
    |-  (pro4.west);
\end{tikzpicture}
\caption{The flow chart of M/X vs B classification, discussed in Section~\ref{class_models}. After inputting all training samples with class M/X and B into the trained LSTM model, we use the output $\hat{I}$ together with $I$ to decide an optimal threshold between M/X and B with trisection method. The loop time is set to 5.}
\label{M/X_b_alg}
\end{figure}

\subsection{Extension to Classification Models} \label{class_models} 

In this section, we introduce \textit{binary classification models} that are built upon the mixed LSTM regression model in Section~\ref{model_description}. The binary classification models are designed for classifications of M/X versus B, M/X versus B/Q and M/X versus C/B/Q.
 
 For M/X versus B, i.e. strong/weak flare classification, we only consider training samples that have flare intensities ranging from $[-7,-6)\cup[-5,-3)$. Borrowing the idea from transfer learning in \cite{NIPS2014_5347}, we make use of the output given by the mixed LSTM regression model, $\hat{I}$, to decide an optimal threshold between M/X and B flares.

Since we know the observed intensity, $I$ of all training samples, for each potential threshold (${\rm thre}\in(-6,-5]$) for $\hat{I}$, we can construct a confusion matrix, where true positives TP = $\sum \mathbbm{1}(\hat{I_i} \geq {\rm thre}, I_i \geq -5.5)$, false positives FP = $\sum \mathbbm{1}(\hat{I_i} \geq {\rm thre}, I_i < -5.5)$, false negatives FN = $\sum \mathbbm{1}(\hat{I_i} < {\rm thre}, I_i \geq -5.5)$, and true negatives TN = $\sum \mathbbm{1}(\hat{I_i} < {\rm thre}, I_i < -5.5)$, where each term is summed over all available training samples. Then we can calculate the $\text{HSS}_2$ score correspondingly (see \citet{Bobra_2015} for the definition of $\text{HSS}_2$). Again, $\mathbbm{1}(\cdot)$ is an indicator function. Note that, in this case, $I$ only takes values in $[-7,-6)\cup[-5,-3).$ Any number between -6 and -5 could act as the threshold for observed intensity, $I$. We hereby take the value of -5.5.

Next we apply the trisection method \citep{doi:10.1137/05063060X} to find the threshold that yields the highest $\text{HSS}_2$. For each iteration, we obtain a ${\rm thre}_{\rm lo}$ and a ${\rm thre}_{\rm up}$ by trisecting the current range of threshold. By constructing confusion matrixs respectively, we compare the $\text{HSS}_2$ score, choose the one with the higher score, and define new  ${\rm thre}_{\rm lo}$ and ${\rm thre}_{\rm up}$. Throughout the iterations, the range of possible thresholds keeps getting smaller and finally we reach an optimal threshold for $\hat{I}$. The flowchart of the algorithm is in Fig.~\ref{M/X_b_alg}.

The M/X versus B/Q classification model adopts the same strategy as the M/X versus B classification model does on determining the threshold between M/X and B/Q. Different from the M/X versus B/Q and M/X versus B models, the M/X versus C/B/Q classification model no longer has the sweet $[-6,-5)$ buffering area for us to train a threshold. Once we include C flares in the model, the threshold is fixed at $-5$.

We use the following $6$ metrics to evaluate all our binary classifiers: Recall, Precision, the $F_1$ score, the Heidke skill scores ($\text{HSS}_1$, $\text{HSS}_2$), see \citet{Bobra_2015} for the definition of $\text{HSS}_1$ and $\text{HSS}_2$, and the true skill statistics (TSS), among which $\text{HSS}_2$ and TSS are our main focuses.  {Specifically, Recall and Precision are two standard metrics evaluating the quality of a prediction. The $F_1$ score is the harmonic mean of Recall and Precision. However, these three scores can be rather unstable when encountering unbalanced samples; which is true in our case where the B/C flares outnumber the M/X flares. We consider TSS and $\text{HSS}_2$ as two reasonable measures of classification performance for solar flares. TSS is invariant to the frequency of samples, unlike Recall or Precision. $\text{HSS}_2$ measures the fractional improvement of the forecast over the random forecast. There are detailed descriptions of $\text{HSS}_1$, $\text{HSS}_2$ and TSS in \citet{Florios_2018}. \cite{Bloomfield_2012} gives conceptual comparison and discussion on the suitability of these metrics when predicting solar flares. A summary of the binary classification results is shown in Section~\ref{class_models_result}.}

\subsection{Test Samples Preparation}  \label{test_sam_prep}

In this paper, we adopt the following strategy for preparing the testing samples to give a fair evaluation of the performance of our algorithms. Recall that each sample is a time series of SHARP parameters and corresponds to a 2-d response variable $[Q,I]$.

First, we take all the samples from the full data set after 2015 (see how we get full data set and do training/testing splitting in Section~\ref{data_preprocess}). For each sample with corresponding response variable $Q=1$ (non-quiet samples), there should be at least one flare happening in the 24-hour time window and the maximum intensity of all the applicable flares should be equal to $I$. For samples with overlapping predictors and the corresponding response variables belonging to the same flare class, we keep one of them at random to avoid repeated predictors - response variable pairs in the testing set. Quiet samples are collected with the same strategy. Section~\ref{results}, \ref{tab:mse}, \ref{class_models_table}, and \ref{q_result_summ} give results for using testing samples obtained via this strategy. 

\section{Results} \label{results}
 In this section, we present results in Sections \ref{reg_result_sec}, \ref{class_models_result} and \ref{quiet_sample_sec} based on the models described in Section~\ref{sec:methodology}. In Section~\ref{useful_info_sec}, we illustrate that under the LSTM architecture, the most efficient time range for predicting the solar activity using the SHARP parameters is within 24 hours before the prediction time. Finally, case studies of intensity prediction with several representative HARPs are given in Section~\ref{case_study_sec}.


With the current time point specified as time 0, we denote a model as ``$[-m,0]$-$[n, n+24]$'' if it uses data in time range $[-m, 0]$ to predict maximum local flare intensities within the $[n, n+24]$ time window ($n,m \geq0$). We define the $[n, n+24]$ time window as \textit{prediction window} and $[-m, 0]$ time window as \textit{input window}. For example, if we want to use the past 6 hours of data to predict the maximum local flare intensity in the 24-hour window $[0, 24]$, the model is denoted as $[-6,0]$-$[0, 24]$. The prediction window is $[0, 24]$ and the input window is $[-6, 0]$ in this case. Similarly, if we want to use the past 12 hours of data to predict the maximum local flare intensity in the next $[12, 36]$ hours, the model should be denoted as $[-12,0]$-$[12, 36]$. The prediction window is $[12, 36]$ and the input window is $[-12, 0]$. 


To allow fair comparisons across models, models with the same prediction window but different input windows are applied to the same group of samples. Consider a series of models: $[-6,0]$-$[0, 24]$, $[-12,0]$-$[0, 24]$, $[-24,0]$-$[0, 24]$ as an example. Their samples are all filtered based on the standard for model $[-24,0]$-$[0, 24]$ (see Section~\ref{pipeline} for details on sample preparation). Therefore, for each sample, we have 24-hour length of SHARP parameters as the predictors; while we only use the last 6 and 12 hours of predictors for models $[-6,0]$-$[0, 24]$ and $[-12,0]$-$[0, 24]$. 

\begin{figure}[htb]
    \centering
    \includegraphics[width=0.9\textwidth]{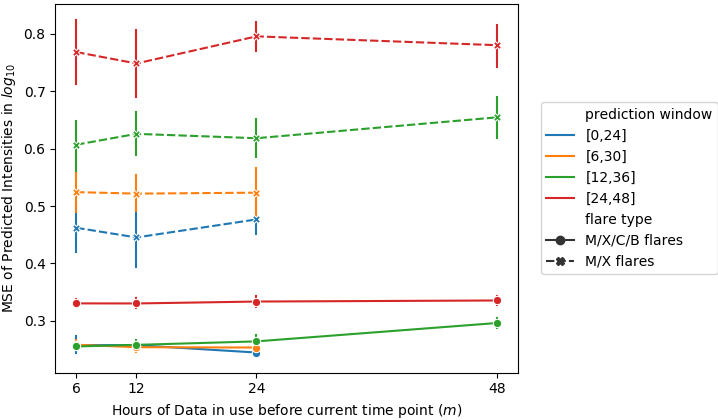}
    \caption{Line chart showing the MSEs of all mixed LSTM regression models, shown in Section~\ref{reg_result_sec}. Again, $m$ is the number of hours of data we use before the current time point, $[-m,0]$ is the input window and $[n,n$+$24]$ is the prediction window. Each point with a vertical line is the average MSE and its 95\% confidence interval of ten regression models with same $[-m,0]$-$[n,n$+$24]$ trained separately. Each line shows the variation of MSE for models with the same prediction window and different lengths of input windows. The solid lines represent the MSEs of all non-quiet testing samples (M/X/C/B). The dashed lines represent the MSEs of those testing samples with M/X flare intensities.}
    \label{fig:visual_mse}
\end{figure}

\subsection{The MSEs from the Mixed LSTM Regression Model} \label{reg_result_sec}

In this section, we present the MSEs of predicted $\log_{10}$ flare intensities from all models in the of line charts. The complete MSE tables for all models and all classes of flares can be found in \ref{tab:mse}.

Fig.~\ref{fig:visual_mse} is a line chart showing the MSEs for models with the same prediction window as the length of input window ($m$) increases (solid lines). The chart also includes the MSEs of the samples with M/X flares (dashed lines). As the prediction window gets farther away from the current time point ($n$ increases), the MSE of all flare samples does not change too much. However, this is not true when we look at MSE calculated from M/X flares only. This shows the sensitivity of the evaluation metric, MSE, with respect to the samples that we use to calculate with. Therefore, the MSE of M/X flares can be considered as another metric for evaluating the performance of the regression models.

Intuitively, the smaller the $n$, i.e., the closer the prediction window from the current time point, the smaller the MSE will be. This is confirmed in Fig.~\ref{fig:visual_mse}. Generally, from the results, the MSE is kept under 0.3 when the prediction window is $[0,24]$, $[6,30]$ or $[12,36]$. We can keep the MSE of M/X flares under 0.5 when $n=0$, a.k.a prediction window is $[0,24]$. We also observe that there is a sudden increase in terms of the MSE of M/X flares when the prediction window is shifted from $[6,30]$ to $[12,36]$ and $[24,48]$. However, we do not observe any significant patterns of the MSE varying monotonically as a function of $m$, the length of the time series that we use for prediction. We elaborate discussions on these results in Section~\ref{useful_info_sec}.



\subsection{Performance of the Classification Models} \label{class_models_result}

We use the $\text{HSS}_2$ score to compare the performances of M/X versus B and M/X versus C/B/Q classifiers. Results in other metrics mentioned in Section~\ref{class_models} are shown in \ref{class_models_table}. In addition, since M/X versus B/Q models give us similar $\text{HSS}_2$ scores as M/X versus B models do, we also put results of M/X versus B/Q models in \ref{class_models_table}.

\begin{figure}[htb]
    \centering
    \includegraphics[width=0.9\textwidth]{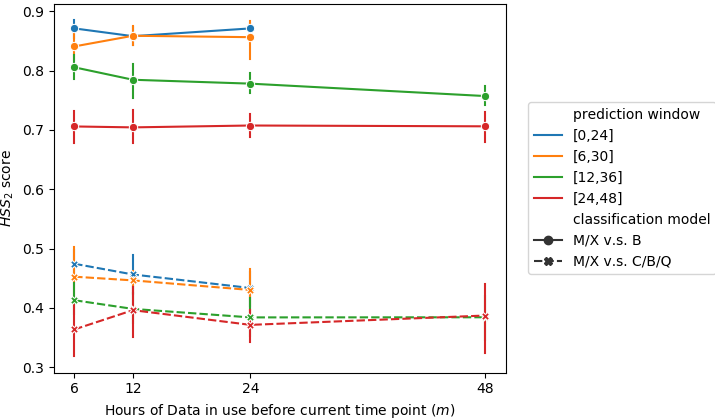}
    \caption{Line chart showing the $\text{HSS}_2$ scores of all classification models, covered in Section~\ref{class_models_result}. Similar to Fig.~\ref{fig:visual_mse}, each point with a vertical line is the average $\text{HSS}_2$ and its 95\% confidence interval of ten classification models with same $[-m,0]$-$[n,n$+$24]$ trained separately. Each line shows the variation of $\text{HSS}_2$ for models with same prediction window and different length of input windows. The solid lines represent the $\text{HSS}_2$s of M/X versus B models. The dashed lines represent the $\text{HSS}_2$ scores of M/X versus C/B/Q models.}
    \label{fig:visual_hss}
\end{figure}

The $\text{HSS}_2$ score results are also shown in the form of a line chart in Fig.~\ref{fig:visual_hss}. There is a large gap between all M/X versus B models and all M/X versus C/B/Q models. As mentioned in Section~\ref{class_models}, we have a intensity interval, $[-6, -5)$ (for C flares), where there are no flares defined as M/X or B. This is mainly why we can get incredibly high scores ($\text{HSS}_2 > 0.8$ when the prediction window is $[0,24]$ or $[6,30]$, $\text{HSS}_2 > 0.7$ when all models) for M/X versus B. As for the M/X versus C/B/Q model, we can hardly get $\text{HSS}_2$ scores greater than 0.5. We manage to classify roughly half of the M and X flares out of other flares when prediction window is $[0,24]$ (See \ref{class_models_table}). Almost all of the mis-classified M and X flares have predicted intensities falling into C flares' intensity range (See Fig.~\ref{fig:dist_predicted}). We do not observe an obvious $\text{HSS}_2$ score difference between models with prediction window $[0,24]$ and $[6,30]$. 
But when the prediction window is shifted from $[6,30]$ to $[12,36]$ and $[24,48]$, there is a large decrease in terms of the $\text{HSS}_2$ score.
\subsection{Results of Quiet Samples from the Mixed LSTM Regression Model} \label{quiet_sample_sec}

\begin{figure}[htb]
    \centering
    \includegraphics[width=0.9\textwidth]{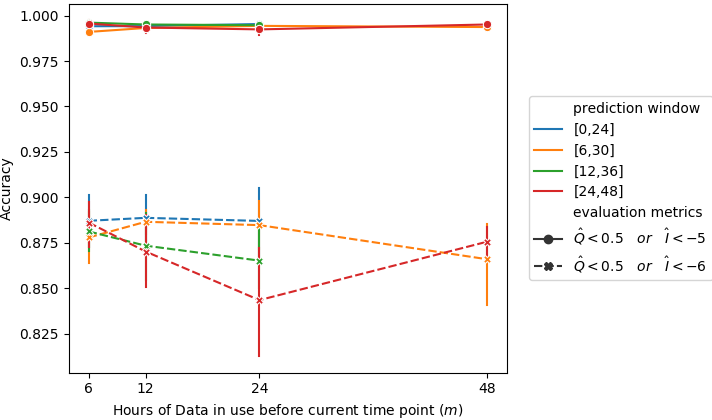}
    \caption{Line chart showing the classification accuracy of quiet samples in all models, covered in Section~\ref{reg_result_sec}. Each point with a vertical line is the average accuracy and its 95\% confidence interval of ten models with same $[-m,0]$-$[n,n$+$24]$ trained separately. Each line shows the variation of the accuracy for models with same prediction window and different length of input windows. The solid lines represent the accuracy when the evaluation metric is $\hat{Q}<0.5$ or $\hat{I}<6$. The dashed lines represent the accuracy when the evaluation metric is $\hat{Q}<0.5$ or $\hat{I}<5$.}
    \label{fig:q_result}
\end{figure}
 
In Section~\ref{reg_result_sec} and Section~\ref{class_models_result}, we only summarise the prediction results of non-quiet samples, i.e. samples with response variables $Q=1$. In this section, we will particularly focus on the performance of all the models in terms of the quiet samples, i.e. samples with response variables $Q=0$.
 
First, we examine the fitted distribution of the predicted intensity ($\hat{I}$) of the quiet samples in Fig.~\ref{fig:dist_predicted}. This is an example of a [-6,0]-[0,24] model. We observe that almost all of the quiet samples have $\hat{I}<-5$ in the testing set, which indicates that the false alarm (False Positive rate) of quiet samples can be restrained significantly in our models. Next, we formally evaluate the performance of the prediction. Note that we don't have the exact observed intensity ($I=N/A$) for quiet samples (see examples of how we define response variables in Table~\ref{example_trans}). Therefore, we consider the prediction result ($[\hat{Q},\hat{I}]$) as successful if it meets either of the following two requirements: (1) the predicted intensity $\hat{I}<k$, (2) predicted quiet score $\hat{Q}<0.5$. Specifically, $k$ takes the value of -5 and -6, where $k=-5$ evaluates the rate of falsely predicting a quiet sample as intensive flare (M and X flare) while $k=-6$ evaluates the rate of falsely predicting a quiet sample as M, X or C flare. We denote $k=-5$ as metric 1 and  $k=-6$ as metric 2. 
 
Fig.~\ref{fig:q_result} shows the summarised result of the quiet sample prediction, where solid line corresponds to metric 1 and dashed line to metric 2 (the summary table can be seen in ~\ref{q_result_summ}). We obtain an accuracy of over 98.5\% for all models in terms of metric 1 and over 80\% in terms of metric 2. Recall that ``-5'' is the cutoff of the logarithm of flare intensity for B and C flares, thus as long as we don't give a $\hat{I}>-5$ which is an alarm of intense flare, we can consider the prediction satisfying. Therefore, we conclude that our regression models have an excellent performance on restraining false alarms.

\begin{figure}[htb]
    \centering
    \includegraphics[width=0.92\textwidth]{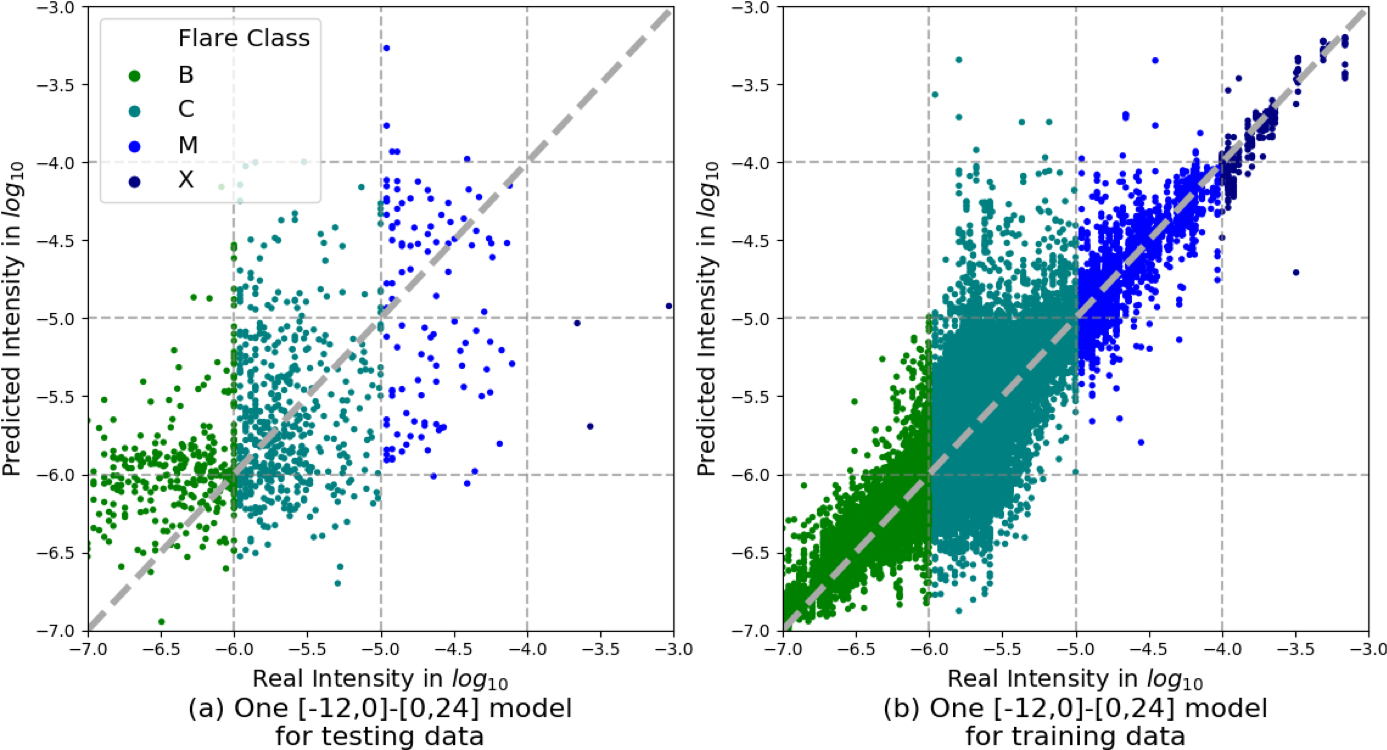}
    \caption{Predicted intensities versus True intensities. Each point represents a recorded flare. Purple stands for X flare, blue for M, aqua for C and green for B. For each panel, the x-axis is the observed intensity and y-axis is the predicted intensity. The thick gray dashed line $y$=$x$ shows the ideal positions where every point should locate when being accurately predicted.}%
    \label{fig:visual_612}%
\end{figure}

\subsection{Post-hoc Analysis} \label{useful_info_sec}

In this section, we show visualizations of the prediction results, combined with the regression and classification results shown in Section~\ref{reg_result_sec}, \ref{class_models_result} and \ref{quiet_sample_sec}, to investigate in-depth how the information in the data (time series of SHARP parameters) convey for solar flare predictions under the LSTM architecture.

\begin{figure}[htb]
    \centering
    \includegraphics[width=0.92\textwidth]{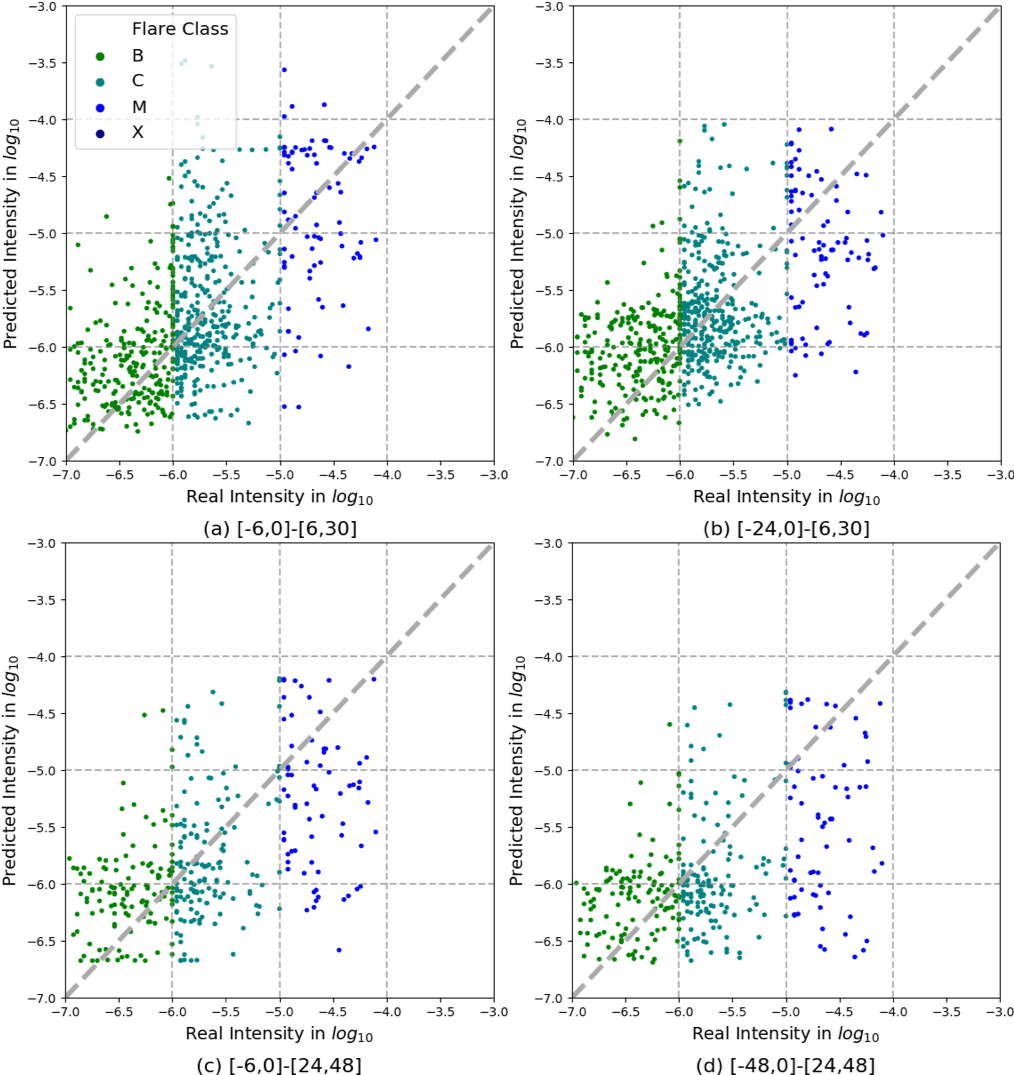}
    \caption{Visualizations for 4 example models. The figures share the same setting as Fig.~\ref{fig:visual_612}. Noted that, in both Fig.~\ref{fig:visual_example} and \ref{fig:dist_predicted}, there are no X flare plotted. Recall that we define the prediction window as $[n, n$+$24]$. Generally, there is no applicable X flares in testing set for $n$ $>$ 0. We have very few X flares. Most of them happened before 2015. For the limited X flares happened after 2015, they either have many frames missing before it happened, or happened only few hours after the video starting. So we don't have X flares in testing set for models with prediction windows farther away from the current time point.}
    \label{fig:visual_example}
\end{figure}

Fig.~\ref{fig:visual_612} and Fig.~\ref{fig:visual_example} show the predicted intensity against the observed intensity with each point representing a flare event. Each color in the figures represents one class of solar flare. Purple stands for X flare, blue for M, aqua for C and green for B. Specifically, except that Fig.~\ref{fig:visual_612}(b) is plotted based on the training samples, all other sub-figures in Fig.~\ref{fig:visual_612} and Fig.~\ref{fig:visual_example} are plotted based on testing samples corresponding to 5 models with different prediction windows and input windows.  {Fig.~\ref{fig:visual_612}(b) exhibits the best performance over all figures, since it is based on training set. We cannot expect to achieve this high accuracy when applying models to the testing set.}

\begin{figure}[ht]
    \centering
'    \includegraphics[width=\textwidth]{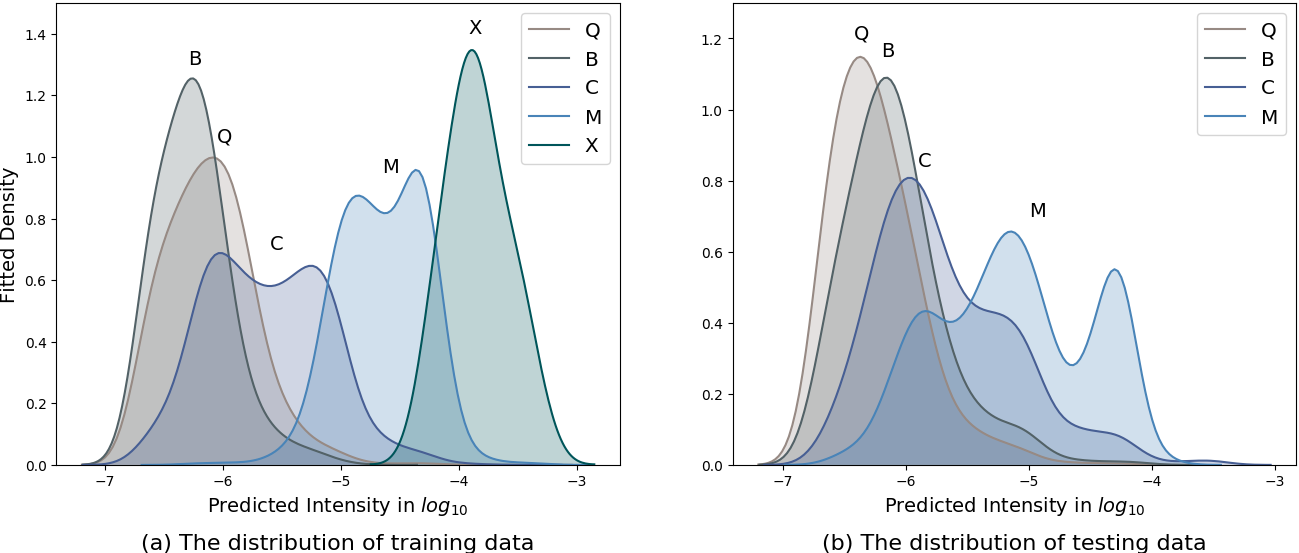}
    \caption{Fitted distribution of predicted intensities based on one $[-6,0]$-$[0,24]$ model. The distribution is fitted using Gaussian kernel with bandwidth=0.15. X-axis is the values taken by predicted intensities, Y-axis stands for the density of fitted distribution. Ideally, flares with class B, C or M should follow an asymptotically normal distribution. The predicted distribution (a) for training data is close to the ideal setting. While for testing set (b), the predicted intensities are still having a hard time separating themselves with other flares.}
    \label{fig:dist_predicted}
\end{figure}

Fig.~\ref{fig:dist_predicted} shows the fitted Gaussian distribution of each class's predicted intensity. The left panel is the fitted Gaussian distribution for training samples and right is for testing samples. Each color represents one class of flares. It can be seen that the different classes of flares, especially neighboring ones, have overlapping predicted intensity values. Nevertheless, the strong flares and weak flares (or quiet time) are still highly distinctive.

Not surprisingly, the farther the prediction window from the current time point, the worse the prediction results. This is also intuitive: predicting what happens after one hour is easier than predicting what happens after ten hours. Another finding is that considering more data backwards (greater $m$) does not necessarily guarantee a better prediction result. The explanation is twofold. 

First, we speculate that the most useful information for predicting the behavior of the prediction window is within 24 hours beforehand. Here 24 denotes the hours from the center of the prediction window to now. Once $n+12\geq24$ (12 is half of the prediction window's length), considering more information does not help much based on our results. Notice that, even though the TSS and $\text{HSS}_2$ scores decrease as the $n$ increases, they always experience a sharp drop when the prediction windows move farther away from $[6,30]$ to $[12,36]$, i.e. $n$ increases from 6 to 12 in all models. Recall that $k$ in Fig.~\ref{lstm_alg} is the number of frame(s) we kept after going through LSTM layers and we take $k=1$ for all our models. Therefore, we are essentially using the output information of the last frame ($n$ hours from the prediction window) to predict the behavior in the prediction window. A worse result indicates that the last frame is less relevant to the prediction window or it is harder for LSTM to build a relationship between the prediction window and the last frame. Thus, the sharp drop when the prediction window shifts from [6,30] to [12,36] indicates the solar activities within the 24 hour window prior to the events have a significant influence on the behavior in the prediction window.

Second, even though the most useful information for prediction is within 24 hours before the events, considering more information offering us worse result is still counter-intuitive. This is due to the limitations of the LSTM model. The LSTM is an artificial recurrent neural network (RNN) architecture used for digging out the temporal properties within time-series data. The parameter matrices for each gate remain unchanged for all input time series. Therefore, the LSTM considers the entire time evolution process in a homogeneous way. If the whole time series before the event is not acting homogeneously, adding information 24 hours before can, on the contrary, impair the performance of the prediction.

\begin{figure}[htb]
    \centering
    \includegraphics[width=0.99\textwidth]{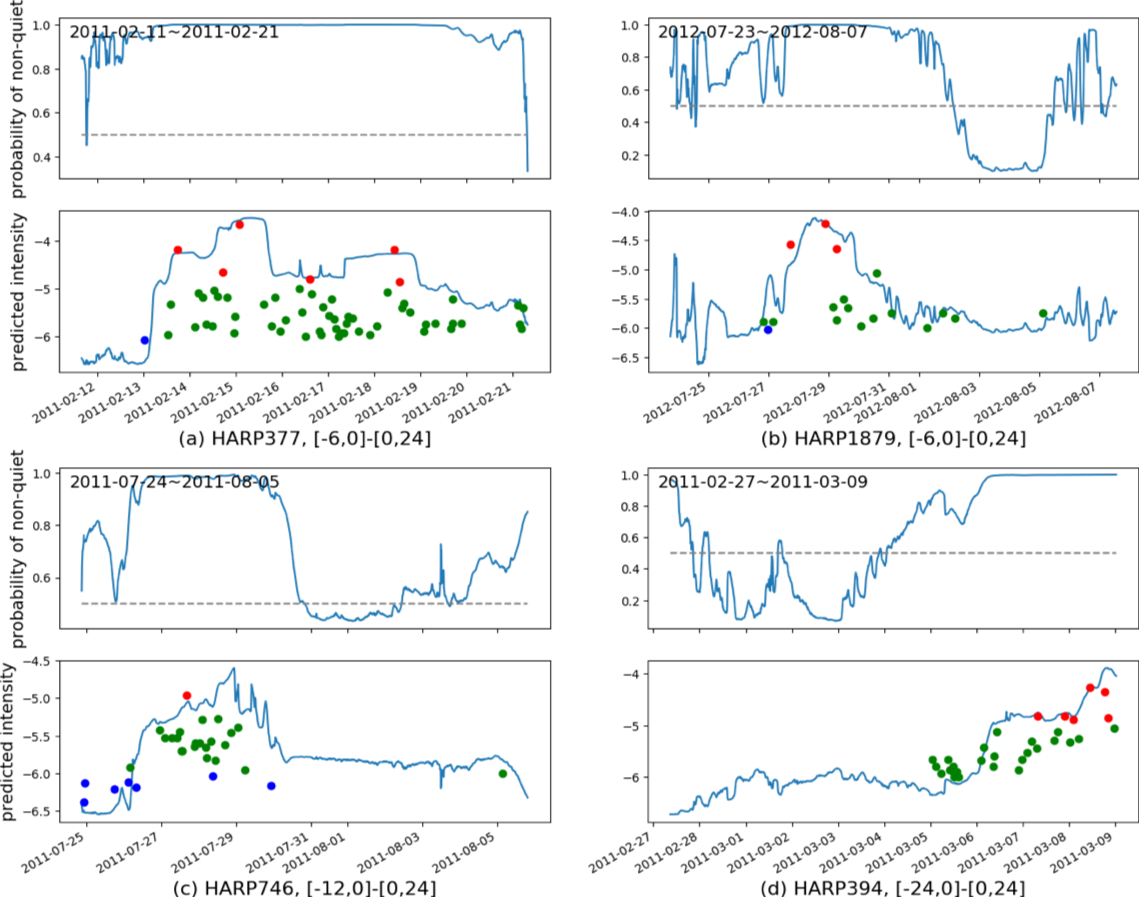}
    \caption{Case Studies: Successful cases. For each plot, the blue curve on the upper panel is the predicted $\hat{Q}$ score. The grey dashed line taking the value of 0.5 is the threshold of dividing quiet and non-quiet times. The blue curve on the lower panel is the predicted real-time flare intensity, $\hat{I}$. There is no time shift on each plot. Each red, green or blue round point corresponds to one recorded M/X, C or B flare respectively. Unlike Fig.~\ref{fig:exampleofar}, the height of each point is exactly the $\log_{10}$ intensity of the flare it represents.}
    \label{fig:case_study_success}
\end{figure}
        

\begin{figure}[htb]
    \centering
    \includegraphics[width=\textwidth]{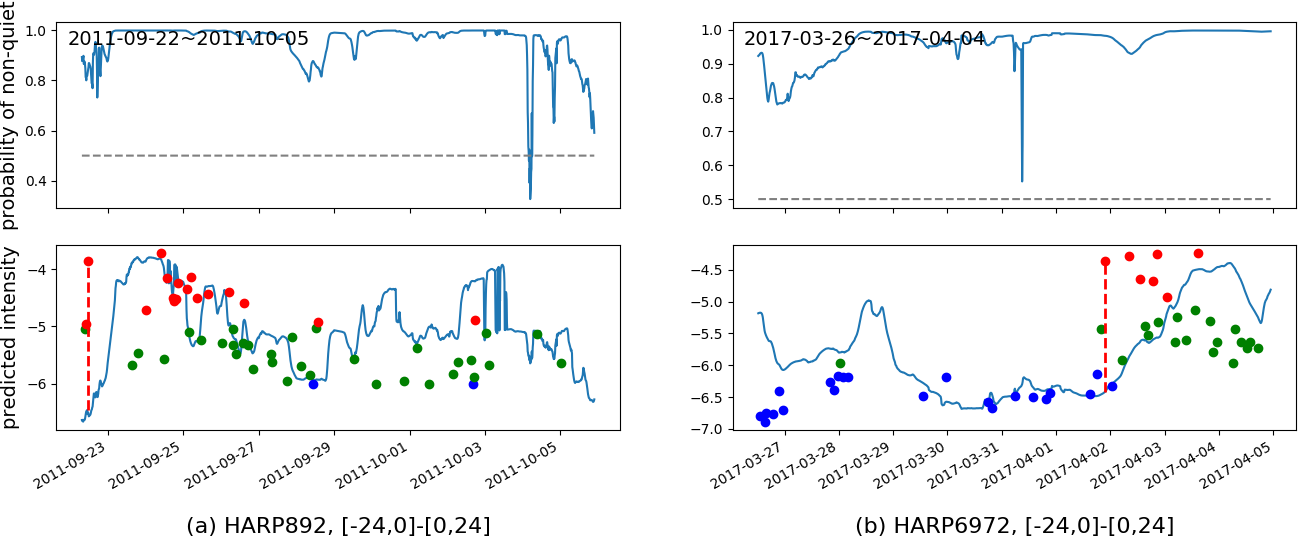}
    \caption{Case Studies: Failed cases. Same setting as Fig.~\ref{fig:case_study_success}. In addition, the red vertical dashed line is to indicate the largest prediction error.}
    \label{fig:case_study_fail}
\end{figure}

\subsection{Case Study} \label{case_study_sec}

In the case study section, we focus on the model performances on M and X flares' predictions for two reasons. First, M and X flares are of primary concern in the flare prediction problem. Second, as shown in Fig.~\ref{fig:visual_mse}, the model can already offer us a decent prediction, i.e. a relatively small MSE, for B and C flares. Besides, Fig.~\ref{fig:dist_predicted} shows that, for both the training and testing set, quiet samples' predicted intensities are restricted below -5. Hence, M and X flares are not only the most important but also the most difficult flares to predict, i.e. generating the highest MSE.

Fig.~\ref{fig:case_study_success} and Fig.~\ref{fig:case_study_fail} show 6 prediction plots, including 4 well and 2 badly performed examples, each of which corresponds to one HARP and one model. The 4 well-performed examples in Fig.~\ref{fig:case_study_success} are chosen where at least one of their M and X flares lays near the $y=x$ diagonal line in Fig.~\ref{fig:visual_example}(a) and (b). For the 2 badly-performed cases in Fig.~\ref{fig:case_study_fail}, we choose two videos where one of their M or X flares has the largest prediction error ($|I-\hat{I}|$) among all M and X flares in the training set and testing set respectively in a $[-24,0]$-$[0,24]$ model.

A successful case should have the blue curve in the lower panel of each plot locating as close as possible to the local maximum flare, i.e. local highest round point. Noted that the existence of dimension $Q$ in the response variable is only to compensate for the non-observable flares. Thus, the quiet score $\hat{Q}$ in the upper panel is more than a signal instead of an exact prediction result. As long as the lower panel offers a $\hat{I} \leq -6$, we can still consider the model having a good prediction of the quiet time.
 
The two cases shown in Fig.~\ref{fig:case_study_fail} represent two typical situations where M and X are wrongly predicted. (1) The model does perceive the increase in flare intensity but not precisely, like in Fig.~\ref{fig:case_study_fail}(a). Predicted intensity may have increased hours before or after the intensive flares' happening. (2) The model fails to detect the intense flares totally, like in Fig.~\ref{fig:case_study_fail}(b). However, this scenario only happens when the certain M/X flares lay at the head or tail of the video. Moreover, videos also tend to have a few frames missing at the beginning and the end. Thus  {we speculate that} it is the potential problem of the  {missing frames} and  {the mismatch of HARP and Active regions (see Section~\ref{response_variable} for details)} rather than the model that restricts the performance of the prediction.  {We also note that there are many missing B and C flares in the GOES data set, which might reduce precision of the response variable, leading to biased prediction results. }

\section{Summary and Discussion} \label{discussion}

In this paper, we presented a pipeline to prepare and analyze data from the SHARP parameters and GOES data set. An LSTM mixed regression model was introduced and we shared encouraging results on solar flare intensity prediction and classifications. The work in this article can be considered as a first step in early prediction of intense solar flare events. Compared to our previous results in \citet{doi:10.1029/2019SW002214}, the models presented in this paper stand out in several aspects.
\begin{itemize}
    \item The prediction score, TSS and $\text{HSS}_2$ of M/X versus B is increased by 0.1 when the prediction window is $[0,24]$.
    \item We predict the exact intensity rather than the class of the flares. 
    \item We consider more cases, including $[-6,0]$-$[0,24]$, $[-12,0]$-$[0,24]$, $[-24,0]$-$[0,24]$; $[-6,0]$-$[6,30]$, $[-12,0]$-$[6,30]$, $[-24,0]$-$[6,30]$; $[-6,0]$-$[12,36]$, $[-12,0]$-$[12,36]$, $[-24,0]$-$[12,36]$, $[-48,0]$-$[12,36]$; $[-6,0]$-$[24,48]$, $[-12,0]$-$[24,48]$, $[-24,0]$-$[24,48]$, $[-48,0]$-$[24,48]$ and prepare the data to offer fair comparison with same prediction windows.
\end{itemize}


There are several promising areas for future work. First, the Sun's activity level experiences an 11 year cycle, the Sun being in the $24^{\rm th}$ cycle that began in December 2008 (\citet{Swpc.noaa.gov}). The boundary between the training and testing sets in this paper are set at year 2015. Flares events that happened after 2015 are not exactly equivalent or comparable to flares before 2015. It would be worthwhile to explore other splits of the data sets into training and testing subsets. Second, in our models, we consider videos of different HARPs equally, which is certainly not the case due to the intrinsic variability among different HARPs. Moreover, there is a latent dependency among flares in the same HARP, which are not modeled in our LSTM approach. Last, as mentioned in Section~\ref{case_study_sec}, our results are limited by its sole dependency on the SHARP parameters, which may or may not fully capture the information of the magnetic field. In the future, we plan to directly work with the HMI magnetograms for real time prediction of flares.  
\appendix

\section{MSE Table for Mixed LSTM Regression} \label{tab:mse}

In this table and all the following tables in the appendix, we denote the $[-m,0]$-$[n,n+24]$ model as $(n+12)$-$m$ for simplicity. For example, $[-12,0]$-$[0,24]$ is 12-12 and $[-24,0]$-$[24,48]$ is 36-24. Note that the values given in the table are based on $\log_{10}$ scale of flare intensity values.

\begin{table}[ht]
\centering
\begin{tabular}{|c|cccccccc|}
\rowcolor[HTML]{C0C0C0} 
Class     & \multicolumn{8}{c|}{\cellcolor[HTML]{C0C0C0}Num of hours before Event - Num of hours of data used} \\
\rowcolor[HTML]{C0C0C0} 
              & 12-06      & 12-12      & 12-24          & 24-12     & 24-24     & 24-48     & 36-06 & 36-24\\ 
\rowcolor[HTML]{CBCEFB} 
Average        & 0.25       & 0.25       & 0.24            & 0.25      & 0.27      & 0.28      &0.29 & 0.30\\
\rowcolor[HTML]{ECF4FF} 
M/X            & 0.44       & 0.46       & 0.48            & 0.61      & 0.63      & 0.69      &0.72 & 0.71\\
\rowcolor[HTML]{CBCEFB} 
C            & 0.19       & 0.20       & 0.19            & 0.14      & 0.19      & 0.16  &0.15 &    0.15\\
\rowcolor[HTML]{ECF4FF} 
B       & 0.25       & 0.23       & 0.22           & 0.29      & 0.25      & 0.27   &0.26 &0.28   \\ 
\end{tabular}
\end{table}

\section{Tables of Classification Results}
\label{class_models_table}

\subsection{M/X versus B flare classification results (calculated based on Table~\ref{mx_b Confusion Matrices})}
\begin{table}[H]
\centering
\begin{tabular}{|c|cccccccc|}
\cellcolor[HTML]{C0C0C0}Metrics                         & \multicolumn{8}{c|}{\cellcolor[HTML]{C0C0C0}Num of hours before Event - Num of hours of data used}  \\ \cline{2-9} 
\multicolumn{1}{|c|}{\cellcolor[HTML]{C0C0C0}} & \multicolumn{1}{c}{\cellcolor[HTML]{C0C0C0}12-06} & \multicolumn{1}{c}{\cellcolor[HTML]{C0C0C0}12-12} & \multicolumn{1}{c}{\cellcolor[HTML]{C0C0C0}12-24} &  \multicolumn{1}{c}{\cellcolor[HTML]{C0C0C0}24-12} & \multicolumn{1}{c}{\cellcolor[HTML]{C0C0C0}24-24} &
\multicolumn{1}{c}{\cellcolor[HTML]{C0C0C0}24-48} &
\multicolumn{1}{c}{\cellcolor[HTML]{C0C0C0}36-06} &
\multicolumn{1}{c|}{\cellcolor[HTML]{C0C0C0}36-24} \\ 
\rowcolor[HTML]{CBCEFB} 
Recall         & 0.89  & 0.89  & 0.91  & 0.80  & 0.80  & 0.80 & 0.74 & 0.74\\
\rowcolor[HTML]{ECF4FF} 
Precision      & 0.92  & 0.92  & 0.93  & 0.89  & 0.92  & 0.91 & 0.94 & 0.94\\
\rowcolor[HTML]{CBCEFB} 
$F_1$ Score    & 0.91  & 0.91  & 0.92  & 0.85  & 0.85  & 0.85 & 0.82 & 0.82\\
\rowcolor[HTML]{ECF4FF} 
$\text{HSS}_1$ & 0.82  & 0.81  & 0.84  & 0.71  & 0.72  & 0.72 & 0.68 & 0.69\\
\rowcolor[HTML]{CBCEFB} 
$\text{HSS}_2$ & 0.86  & 0.86  & 0.88  & 0.75  & 0.78  & 0.76 & 0.71 & 0.71\\
\rowcolor[HTML]{ECF4FF} 
TSS            & 0.85  & 0.85  & 0.88  & 0.74  & 0.76  & 0.75 & 0.69 & 0.70\\ 
\end{tabular}
\label{tab:M/X_vs_b_result}
\end{table}

\subsection{M/X versus B/Q flare classification results (calculated based on Table~\ref{mx_bq Confusion Matrices})}
\begin{table}[H]
\centering
\begin{tabular}{|c|cccccccc|}
\cellcolor[HTML]{C0C0C0}Metrics                         &
\multicolumn{8}{c|}{\cellcolor[HTML]{C0C0C0}Num of hours before Event - Num of hours of data used}  \\ \cline{2-9} 
\multicolumn{1}{|c|}{\cellcolor[HTML]{C0C0C0}} & \multicolumn{1}{c}{\cellcolor[HTML]{C0C0C0}12-06} & \multicolumn{1}{c}{\cellcolor[HTML]{C0C0C0}12-12} & \multicolumn{1}{c}{\cellcolor[HTML]{C0C0C0}12-24} &  \multicolumn{1}{c}{\cellcolor[HTML]{C0C0C0}24-12} & \multicolumn{1}{c}{\cellcolor[HTML]{C0C0C0}24-24} & \multicolumn{1}{c|}{\cellcolor[HTML]{C0C0C0}24-48}&
\multicolumn{1}{c|}{\cellcolor[HTML]{C0C0C0}36-06}& 
\multicolumn{1}{c|}{\cellcolor[HTML]{C0C0C0}36-24} \\ 
\rowcolor[HTML]{CBCEFB} 
Recall         & 0.91  & 0.89  & 0.90  & 0.79  & 0.80  & 0.80  & 0.74  & 0.74\\
\rowcolor[HTML]{ECF4FF} 
Precision      & 0.64  & 0.66  & 0.66  & 0.72  & 0.71  & 0.68  & 0.68  & 0.66\\
\rowcolor[HTML]{CBCEFB} 
$F_1$ Score    & 0.75  & 0.75  & 0.76  & 0.75  & 0.75  & 0.73  & 0.70  & 0.69\\
\rowcolor[HTML]{ECF4FF} 
$\text{HSS}_1$ & 0.39  & 0.42  & 0.43  & 0.48  & 0.46  & 0.39  & 0.34  & 0.31\\
\rowcolor[HTML]{CBCEFB} 
$\text{HSS}_2$ & 0.73  & 0.74  & 0.74  & 0.73  & 0.72  & 0.70  & 0.67  & 0.66\\
\rowcolor[HTML]{ECF4FF} 
TSS            & 0.88  & 0.86  & 0.87  & 0.76  & 0.77  & 0.76  & 0.70  & 0.70\\ 
\end{tabular}
\label{tab:M/X_vs_B/Q_result}
\end{table}

\subsection{M/X versus C/B/Q flare classification results (calculated based on Table~\ref{mx_o Confusion Matrices})}
\begin{table}[H]
\centering
\begin{tabular}{|c|cccccccc|}
\cellcolor[HTML]{C0C0C0}Metrics                         & \multicolumn{8}{c|}{\cellcolor[HTML]{C0C0C0}Num of hours before Event - Num of hours of data used}  \\ \cline{2-9} 
\multicolumn{1}{|c|}{\cellcolor[HTML]{C0C0C0}} & \multicolumn{1}{c}{\cellcolor[HTML]{C0C0C0}12-06} & \multicolumn{1}{c}{\cellcolor[HTML]{C0C0C0}12-12} & \multicolumn{1}{c}{\cellcolor[HTML]{C0C0C0}12-24} & \multicolumn{1}{c}{\cellcolor[HTML]{C0C0C0}24-12} & \multicolumn{1}{c}{\cellcolor[HTML]{C0C0C0}24-24} & \multicolumn{1}{c}{\cellcolor[HTML]{C0C0C0}24-48} & \multicolumn{1}{c}{\cellcolor[HTML]{C0C0C0}36-06} & \multicolumn{1}{c|}{\cellcolor[HTML]{C0C0C0}36-24} \\ 
\rowcolor[HTML]{CBCEFB} 
Recall         & 0.54  & 0.49  & 0.45  & 0.35  & 0.34  & 0.32  & 0.29  & 0.32\\
\rowcolor[HTML]{ECF4FF} 
Precision      & 0.45  & 0.47  & 0.47  & 0.54  & 0.52  & 0.53  & 0.55  & 0.56\\
\rowcolor[HTML]{CBCEFB} 
$F_1$ Score    & 0.49  & 0.48  & 0.46  & 0.42  & 0.41  & 0.40  & 0.38  & 0.40\\
\rowcolor[HTML]{ECF4FF} 
$\text{HSS}_1$ & -0.11  & -0.06  & -0.05  & 0.05  & 0.02  & 0.03  & 0.06  & 0.07\\
\rowcolor[HTML]{CBCEFB} 
$\text{HSS}_2$ & 0.47  & 0.45  & 0.44  & 0.39  & 0.38  & 0.37  & 0.35  & 0.37\\
\rowcolor[HTML]{ECF4FF} 
TSS            & 0.51  & 0.46  & 0.43  & 0.33  & 0.32  & 0.30  & 0.28  & 0.30\\ 
\end{tabular}
\label{tab:M/X_vs_others_result}
\end{table}

\subsection{M/X versus B Confusion Matrices} \label{mx_b Confusion Matrices}

\begin{table}[H]
\centering
\begin{tabular}{|c|cccc|}
\rowcolor[HTML]{C0C0C0} 
Model & \multicolumn{4}{c|}{\cellcolor[HTML]{C0C0C0}Confusion Matrix (mean [min, max])} \\
\rowcolor[HTML]{C0C0C0} 
      & TP                   & FN                & FP              & TN                  \\
\rowcolor[HTML]{ECF4FF} 
12-06 & 86.2 [83,88]         & 8.8 [7,12]        & 7.3 [1,14]      & 176.7 [170,183]     \\
\rowcolor[HTML]{CBCEFB} 
12-12 & 84.2 [80,88]         & 10.8 [7,15]       & 6.8 [3,10]      & 177.2 [174,181]     \\
\rowcolor[HTML]{ECF4FF} 
12-24 & 85.4 [79,88]         & 9.6 [7,16]        & 6.4 [4,8]       & 177.6 [176,180]     \\
\rowcolor[HTML]{CBCEFB} 
18-06 & 79.5 [74,86]         & 10.5 [4,16]      & 7.9 [3,19]       & 156.1 [145,161]     \\
\rowcolor[HTML]{ECF4FF} 
18-12 & 79.2 [76,84]         & 10.8 [6,14]      & 5.4 [1,12]       & 158.6 [152,163]     \\
\rowcolor[HTML]{CBCEFB} 
18-24 & 81.1 [75,88]         & 8.9 [2,15]      & 7.9 [1,35]       & 156.1 [129,163]     \\
\rowcolor[HTML]{ECF4FF} 
24-06 & 71.7 [66,78]         & 17.3 [11,23]      & 4.3 [2,7]       & 158.7 [156,161]     \\
\rowcolor[HTML]{CBCEFB} 
24-12 & 70.3 [63,76]         & 18.7 [13,26]      & 5.2 [1,9]       & 157.8 [154,162]     \\
\rowcolor[HTML]{ECF4FF} 
24-24 & 71.0 [66,76]         & 18.0 [12,23]      & 6.8 [3,12]      & 156.2 [151,160]     \\
\rowcolor[HTML]{CBCEFB} 
24-48 & 64.4 [60,71]         & 16.6 [10,21]      & 6.4 [3,12]      & 113.6 [108,117]     \\
\rowcolor[HTML]{ECF4FF} 
36-06 & 57.5 [49,63]         & 20.5 [15,29]      & 4.1 [2,9]       & 89.9 [85,92]     \\
\rowcolor[HTML]{CBCEFB} 
36-12 & 59.9 [53,67]         & 18.1 [11,25]      & 6.8 [2,17]       & 87.2 [77,92]     \\
\rowcolor[HTML]{ECF4FF} 
36-24 & 57.6 [53,63]         & 20.4 [15,25]      & 4.1 [2,15]       & 89.9 [79,92]     \\
\rowcolor[HTML]{CBCEFB} 
36-48 & 59.4 [49,65]         & 18.6 [13,29]      & 6.1 [2,14]       & 87.9 [80,92]     \\
\end{tabular}
\end{table}

\subsection{M/X versus B/Q Confusion Matrices}  \label{mx_bq Confusion Matrices}
\begin{table}[H]
\centering
\begin{tabular}{|c|cccc|}
\rowcolor[HTML]{C0C0C0} 
Model & \multicolumn{4}{c|}{\cellcolor[HTML]{C0C0C0}Confusion Matrix (mean [min, max])} \\
\rowcolor[HTML]{C0C0C0} 
      & TP                  & FN              & FP               & TN                    \\
\rowcolor[HTML]{ECF4FF} 
12-06 & 86.2 [83,88]        & 8.8 [7,12]      & 49.0 [29,73]     & 1606.0 [1582,1626]    \\
\rowcolor[HTML]{CBCEFB} 
12-12 & 84.2 [80,88]        & 10.8 [7,15]     & 44.3 [33,55]     & 1610.7 [1600,1622]    \\
\rowcolor[HTML]{ECF4FF} 
12-24 & 85.4 [79,88]        & 9.6 [7,16]      & 44.6 [35,57]     & 1610.4 [1598,1620]    \\
\rowcolor[HTML]{CBCEFB} 
18-06 & 79.5 [74,86]        & 10.5 [4,16]      & 63.4 [23,113]     & 1571.6 [1522,1612]    \\
\rowcolor[HTML]{ECF4FF} 
18-12 & 79.2 [76,84]        & 10.8 [6,14]      & 51.3 [27,78]     & 1583.7 [1557,1608]    \\
\rowcolor[HTML]{CBCEFB} 
18-24 & 81.1 [75,88]        & 8.9 [2,15]      & 59.0 [21,167]     & 1576.0 [1468,1614]    \\
\rowcolor[HTML]{ECF4FF} 
24-06 & 71.7 [66,78]        & 17.3 [11,23]    & 25.6 [18,33]     & 915.4 [908,923]       \\
\rowcolor[HTML]{CBCEFB} 
24-12 & 70.3 [63,76]        & 18.7 [13,26]    & 27.8 [14,40]     & 913.2 [901,927]       \\
\rowcolor[HTML]{ECF4FF} 
24-24 & 71.0 [66,76]        & 18.0 [12,23]    & 29.8 [20,40]     & 911.2 [901,921]       \\
\rowcolor[HTML]{CBCEFB} 
24-48 & 64.4 [60,71]        & 16.6 [10,21]    & 32.7 [17,57]     & 865.3 [841,881]       \\
\rowcolor[HTML]{ECF4FF} 
36-06 & 57.5 [49,63]        & 20.5 [15,29]      & 31.1 [8,80]     & 840.9 [792,864]    \\
\rowcolor[HTML]{CBCEFB} 
36-12 & 59.9 [53,67]        & 18.1 [11,25]      & 43.0 [19,99]     & 829.0 [773,853]    \\
\rowcolor[HTML]{ECF4FF} 
36-24 & 57.6 [53,63]        & 20.4 [15,25]      & 33.8 [13,100]     & 838.2 [772,859]    \\
\rowcolor[HTML]{CBCEFB} 
36-48 & 59.4 [49,65]        & 18.6 [13,29]      & 39.8 [21,78]     & 832.2 [794,851]    \\
\end{tabular}
\end{table}

\subsection{M/X versus C/B/Q Confusion Matrices}  \label{mx_o Confusion Matrices}
\begin{table}[H]
\centering
\begin{tabular}{|c|cccc|}
\rowcolor[HTML]{C0C0C0} 
Model & \multicolumn{4}{c|}{\cellcolor[HTML]{C0C0C0}Confusion Matrix (mean [min, max])} \\
\rowcolor[HTML]{C0C0C0} 
      & TP                & FN               & FP               & TN                     \\ 
\rowcolor[HTML]{ECF4FF} 
12-06 & 49.8 [40,57]      & 45.2 [38,55]     & 54.7 [37,67]     & 1998.3 [1986,2016]     \\
\rowcolor[HTML]{CBCEFB} 
12-12 & 47.1 [38,58]      & 47.9 [37,57]     & 53.5 [42,79]     & 1999.5 [1974,2011]     \\
\rowcolor[HTML]{ECF4FF} 
12-24 & 41.6 [32,54]      & 53.4 [41,63]     & 44.7 [31,64]     & 2008.3 [1989,2022]     \\
\rowcolor[HTML]{CBCEFB} 
18-06 & 36.6 [24,51]      & 53.4 [39,66]     & 35.5 [24,54]     & 1856.3 [1838,1868]     \\
\rowcolor[HTML]{ECF4FF} 
18-12 & 37.3 [29,43]      & 52.7 [47,61]     & 31.7 [18,42]     & 1860.3 [1850,1874]     \\
\rowcolor[HTML]{CBCEFB} 
18-24 & 35.0 [26,46]      & 55.0 [44,64]     & 29.2 [16,41]     & 1862.8 [1851,1876]     \\
\rowcolor[HTML]{ECF4FF} 
24-06 & 32.2 [27,40]      & 48.8 [41,54]     & 30.7 [20,38]     & 1137.3 [1130,1148]     \\
\rowcolor[HTML]{CBCEFB} 
24-12 & 29.4 [24,35]      & 51.6 [46,57]     & 26.1 [17,33]     & 1141.9 [1135,1151]     \\
\rowcolor[HTML]{ECF4FF} 
24-24 & 28.8 [19,39]      & 52.2 [42,62]     & 27.7 [20,33]     & 1140.3 [1135,1148]     \\
\rowcolor[HTML]{CBCEFB} 
24-48 & 28.0 [22,38]      & 53 [43,59]       & 25.1 [12,32]     & 1142.9 [1136,1156]     \\
\rowcolor[HTML]{ECF4FF} 
36-06 & 23.6 [12,33]      & 54.4 [45,66]     & 17.0 [10,22]     & 1025.0 [1020,1032]     \\
\rowcolor[HTML]{CBCEFB} 
36-12 & 26.9 [13,36]      & 51.1 [42,65]     & 19.9 [7,33]      & 1022.1 [1009,1035]     \\
\rowcolor[HTML]{ECF4FF} 
36-24 & 25.2 [19,29]      & 52.8 [49,59]     & 21.5 [14,40]     & 1020.5 [1002,1028]     \\
\rowcolor[HTML]{CBCEFB} 
36-48 & 25.1 [9,35]       & 52.9 [43,69]     & 15.9 [9,28]      & 1026.1 [1014,1033]     \\ 
\end{tabular}
\end{table}

\section{Summary of Accuracy of Quiet Sample Prediction} \label{q_result_summ}

\begin{table}[H]
\begin{tabular}{|l|ll|l|ll|}
\rowcolor[HTML]{C0C0C0} 
\multicolumn{1}{|c|}{\cellcolor[HTML]{C0C0C0}Model} & \multicolumn{2}{c|}{\cellcolor[HTML]{C0C0C0}Accuracy (mean [min, max] in \%)}                                 & \multicolumn{1}{c|}{\cellcolor[HTML]{C0C0C0}Model} & \multicolumn{2}{c|}{\cellcolor[HTML]{C0C0C0}Accuracy}                                                        \\
\rowcolor[HTML]{C0C0C0} 
\multicolumn{1}{|c|}{\cellcolor[HTML]{C0C0C0}}      & \multicolumn{1}{c}{\cellcolor[HTML]{C0C0C0}Metric 1} & \multicolumn{1}{c|}{\cellcolor[HTML]{C0C0C0}Metric 2} & \multicolumn{1}{c|}{\cellcolor[HTML]{C0C0C0}}      & \multicolumn{1}{c}{\cellcolor[HTML]{C0C0C0}Metric 1} & \multicolumn{1}{c|}{\cellcolor[HTML]{C0C0C0}Metric 2} \\
\rowcolor[HTML]{ECF4FF} 
12-06                                             & 99.4 [98.9,99.8]                                     & 89.0 [83.5,92.3]                                     & 24-12                                             & 99.6 [99.4,99.9]                                     & 88.2 [85.5,91.3]                                     \\
\rowcolor[HTML]{CBCEFB} 
12-12                                             & 99.4 [98.9,99.8]                                     & 89.1 [86.0,91.6]                                     & 24-24                                             & 99.5 [99.4,99.9]                                     & 87.6 [82.9,91.9]                                     \\
\rowcolor[HTML]{ECF4FF} 
12-24                                             & 99.6 [99.2,99.9]                                     & 88.9 [82.9,92.4]                                     & 24-48                                             & 99.5 [99.2,99.7]                                     & 86.8 [83.4,92.5]                                     \\
\rowcolor[HTML]{CBCEFB} 
18-06                                             & 99.1 [98.7,99.5]                                     & 87.4 [83.8,91.7]                                     & 36-06                                             & 99.6 [99.1,100]                                      & 88.6 [84.8,91.9]                                     \\
\rowcolor[HTML]{ECF4FF} 
18-12                                             & 99.3 [99.0,99.7]                                     & 88.6 [86.9,90.1]                                     & 36-12                                             & 99.4 [98.5,99.9]                                     & 87.5 [82.9,92.5]                                     \\
\rowcolor[HTML]{CBCEFB} 
18-24                                             & 99.4 [99.0,99.7]                                     & 88.8 [85.7,92.7]                                     & 36-24                                             & 99.3 [98.2,99.7]                                     & 84.1 [74.2,90.9]                                     \\
\rowcolor[HTML]{ECF4FF} 
24-06                                             & 99.3 [99.1,99.9]                                     & 86.6 [77.8,90.7]                                     & 36-48                                             & 99.5 [99.0,99.9]                                     & 87.6 [84.1,89.6]                                    
\end{tabular}
\end{table}

\acknowledgments

We thank Professors Gabor Toth, Tuija Pulkkinen, Shasha Zou, and Igor Sokolov from the Climate and Space Sciences and Engineering (CLaSP) at the University of Michigan for helpful comments and discussions. This work is supported by NSF grant AGS-1322543 and NASA grants 80NSSC19K0373 and 80NSSC18K1208. We also acknowledge support from the NASA DRIVE Center at the University of Michigan under grant NASA 80NSSC20K0600. All SHARP data used in this study are available from the Joint Science Operations Center (JSOC) NASA grant, see \url{http://jsoc.stanford.edu/}. All relevant digital values used in the manuscript (both data and model) will be permanently archived at the U-M Library Deep Blue data repository, which is specifically designed for U-M researchers to share their research data and to ensure its long-term viability. Data sets will be assigned Digital Object Identifiers (DOIs) which will serve as identifiers for the data, enabling them to be cited in publications.

\bibliography{flare_intensity_paper}

\end{document}